\documentclass[%
 aip,
 jmp,%
 amsmath,amssymb,
 reprint,%
]{revtex4-2}

\usepackage[version=3]{mhchem} 
\usepackage{balance}
\usepackage{times, mathptmx}
\usepackage{caption}
\usepackage{subcaption}
\usepackage{graphicx}
\usepackage{dcolumn}
\usepackage{bm}

\usepackage[utf8]{inputenc}
\usepackage[T1]{fontenc}
\usepackage{mathptmx}
\usepackage{mathrsfs}
\usepackage{float}
\usepackage{fancyhdr}
\usepackage{fnpos}
\usepackage[english]{babel}
\usepackage{array}
\usepackage{droidsans}
\usepackage{charter}
\usepackage[usenames,dvipsnames]{xcolor}
\usepackage{setspace}
\usepackage[compact]{titlesec}
\usepackage{hyperref}

\begin{document}

\preprint{AIP/123-QED}

\title[]{Effect of Serum Starvation on Rheology of Cell Monolayers}

\author{Abhimanyu Kiran}
 \affiliation{Department of Mechanical Engineering, Indian Institute of Technology, Ropar}%
\author{Chandra Shekhar}%
\affiliation{Department of Chemical Engineering, Indian Institute of Technology, Ropar}%
\author{Manigandan Sabapathy}
\affiliation{Department of Chemical Engineering, Indian Institute of Technology, Ropar}%
\author{Manoranjan Mishra}
\affiliation{Department of Mathematics, Indian Institute of Technology, Ropar}%
\author{Lalit Kumar}
\affiliation{Department of Energy Sciences and Engineering, Indian Institute of Technology, Bombay, India.}%
\author{Navin Kumar}
 \affiliation{Department of Mechanical Engineering, Indian Institute of Technology, Ropar}%
 \author{Vishwajeet Mehandia}
 \email{vishwajeet@iitrpr.ac.in}
\affiliation{Department of Chemical Engineering, Indian Institute of Technology, Ropar}%

\date{\today}

\begin{abstract}
The rheological properties of cells and tissues are central to embryonic development and homoeostasis in adult tissues and organs and are closely
related to their physiological activities. In this work, we present our study of rheological experiments on cell monolayer under serum starvation compared to that of healthy cell monolayer with full serum.
The normal functioning of cells depends on the micronutrient supply provided by the serum in the growth media. Serum starvation is one of the most widely used procedures in cell biology. Serum deficiency may lead 
to genomic instability, variation in protein expression, chronic diseases, and some specific types of cancers. However, the effect of deprivation of serum concentration on the material properties of cells is still
unknown. Therefore, we performed the macro-rheology experiments to investigate the effect of serum starvation on a fully confluent Madin Darby Canine Kidney (MDCK) cell monolayer. The material properties
such as storage modulus ($G^{\prime}$) and loss modulus ($G^{\prime \prime}$), of the monolayer, were measured using oscillatory shear experiments under serum-free (0\% FBS) and full serum (10\% FBS)
conditions. Additionally, the step strain experiments were performed to gain more insights into the viscoelastic properties of the cell monolayer. Our results indicate that without serum, the loss and storage moduli 
decrease and do not recover fully even after small deformation. This is because of the lack of nutrients, which may result in many permanent physiological changes. Whereas, the healthy cell monolayer under full serum condition, remains strong \& flexible, and can fully recover even from a large deformation at higher strain.
\end{abstract}

\maketitle


\section{\label{sec:intro}Introduction}
The mechanical properties of the active gels, cells, and tissues have been an emerging field of research over the past few decades. The deformability is one of the crucial
features of the cells \& tissues that are well-acknowledged but poorly understood \cite{Fregin:2019gx}. The cells are deformed by using micro or macro-rheology techniques, and their mechanical
response as a result of deformation is measured. Most of the studies report the rheology of a single cell using micro-rheology techniques 
\cite{Massiera:2007kv, Sander:2015dg, Emily:2007, HIRATSUKA:2009, LINCOLN:2007, Wottawah:2005fh} or cell monolayers \cite{Bruckner:2017gj, Nehls:2019ep, Bruckner:2018id}. 
The micro-rheology techniques include active force probe optical/magnetic tweezers, passive two-point bead tracking via image-based software, and localized dynamic material deformation through an
atomic force microscope (AFM). On the other hand, the macro-rheology is usually performed using an oscillatory rheometer \cite{Chen:2010jn, Fernandez:2007ch, Dakhil:2016ij}.
The material properties reported by micro-rheology techniques at cellular or subcellular levels underestimate the bulk properties of cells and tissues \cite{Schmidt:2000jz, Weigand:2017eg}
because the cells are highly heterogeneous, so the viscoelastic parameters obtained from micro-rheology experiments are local. On the other hand, the macro-rheology can probe the bulk properties, averaging out
the heterogeneities for one cell and over a large collection of cells in the monolayers \cite{Fernandez:2007ch, Dakhil:2016ij}. 

The living cells are highly complex machines on the planet as they can remodel their cytoskeleton (CSK) to perform various functions like spreading, growth, migration,
division, contraction, and metastasis \cite{Zhou:2013ho}. Hence, instead of structural architecture, the cytoskeleton's dynamic nature fundamentally separates the active cells from passive
non-living materials \cite{Kollmannsberger:2011iy}. Further, the disruption of the cell's CSK dynamics via drugs \cite{Fabry:2001fn, Stamenovic:2004dc, Fabry:2003dr, Bruckner:2017gj}
or mechanical distending \cite{Rosenblatt:2004if, Trepat:2004hs} changes the rheological properties of the cells. As per well-established literature, two fundamental principles that govern the properties of
cells are (1) The pre-existing tension (pre-stress) borne by the cytoskeleton, and (2) The weak power-law response of cells to externally applied stress \cite{Stamenovic:2008ke}.
Ingber's model describes the cell pre-stress \cite{Ingber:2003cp, Ingber:2014fs}, while the soft glass rheology (SGR) model \cite{Lenormand:2004bu, Deng:2006tn, Bursac:2005ex, Trepat:2007jb}
explains the power-law response of the cells. Although both models are quite different from each other; together, they capture the cell's actual response to the applied 
deformation \cite{Rosenblatt:2006gs, Kroy:2007fy}.

  The Cell Monolayer Rheology (CMR) is a well known macro-rheology technique used to investigate the material properties of the cell monolayer. These properties may
vary due to the progression of various diseases like cytotoxicity, malignancy, and other abnormalities. Most of the rheological studies of cells were performed on a discrete cell
monolayer, where a large number of single fibroblasts or HeLa cells are attached to both plates of the rheometer \cite{Sander:2015dg, Fernandez:2007ch, Dakhil:2016ij}. Cell
monolayer rheology can probe a vast number of cells and a better average of mechanical properties. This study uses a similar technique as CMR to probe the cell monolayer
of MDCK epithelial cells. These cells form a continuous cell monolayer, unlike cell monolayer of many disconnected fibroblasts or HeLa cells \cite{Bruckner:2018id}. For the CMR, we have used 
a rheo-microscope (microscope module) of a commercial rheometer.

Further, the micronutrients (proteins and minerals) are essential for regulating cell viability, homeostasis, and DNA metabolic pathways \cite{Pirkmajer:2011fa, Cooper2000}. The
serum is the only source in the culture media to provide the micronutrients to the cells. Generally, the concentration of serum is only 10\% of the complete media \cite{Wrobel:2014dm}.
The lack of micronutrients can cause genomic instability, protein expression variations, disruptions in signalling pathways, and deprivation of growth factors in the cells \cite{Pirkmajer:2011fa, Cooper2000}.
It should also be noted that the effect of serum starvation depends on the protocol and the cell type \cite{Pirkmajer:2011fa, Cooper2000}. With time, the cells consume the micronutrients and their 
concentration decay in the culture media. Recently, \citet{Miyaoka:2011ej} performed micro-rheological experiments using AFM on NIH3T3 fibroblasts under serum starvation. In their study, the
cells were incubated with 0.1\% FBS for 24 hrs and then placed in microarray for another 24 hrs before the AFM study was done. They found that under serum starvation at 0.1\% FBS
the storage modulus $(G^{\prime})$ and loss modulus $(G^{\prime\prime})$ show a weak power-law behaviour with the frequency, but the dependence of loss modulus $(G^{\prime\prime})$ 
on frequency increases after $\le$10 Hz.  However, from the above study, it is utmost important to understand the complete mechanical bulk properties of the cell monolayers with the
effects of serum concentration, which is the main objective of our present paper. 

In this study, we present the bulk rheology of fully confluent epithelial cell monolayer of MDCK cells with full serum at 10\% FBS and without serum at 0\% FBS, using the
Cell Monolayer Rheology (CMR) technique. The cell monolayer was subjected to standard rheological experiments including (1) oscillatory shear which comprises of amplitude and frequency sweep,
and (2) step strain at a specified shear rate to study the recovery of the properties of the cell monolayer. 

Our experimental observations reveal that the cell monolayer's rheological properties depend not only on the applied strain but also on the serum concentration. This work aims to
investigate the macro-rheology of MDCK cell monolayer and determine the effect of serum starvation on its rheological properties.

\section{\label{sec:methods}Material and Methods}
\subsection{\label{sec:cells}Cell Culture}

  We used Madin Darby Canine Kidney (MDCK) II epithelial cells, which are stably transfected with E-Cadherin Green Fluorescent Protein (J. W. Nelson Lab, Stanford University).
We cultured these cells in a growth medium made of 90\% Dulbecco's Modified Eagle Medium (DMEM), low glucose (1\% g/L), pyruvate (Cat No: 11885092, ThemoFisher Scientific)
containing 10\% Fetal Bovine Serum (FBS) and antibiotics at $37^{\circ}$C in a 5\% CO$_{2}$ in a humidified incubator. The cells were replated on a rectangular glass-coverslip (Blue star).
When the cell monolayer became 90-100\% confluent (Fig.~\ref{fig:cmr}), then the growth medium was replaced with Gibco Leibovitz's L15 medium (Cat No: 11415049, ThemoFisher Scientific)
with Fetal Bovine Serum (FBS) and antibiotics. Then these cultures were incubated at $37^{\circ}$C in a 5\% CO$_{2}$ humidified incubator for 8 hours. 

  Serum-free and full serum conditions were created when the cells are incubated with Gibco Leibovitz's L15 medium with 0\% and 10\% FBS at $37^{\circ}$C in a 5\% CO$_{2}$
humidified incubator for 8 hrs, respectively. After this, the cells adhered to the coverslip were used for the rheological experiments Fig.~\ref{fig:expt1}. 

\subsection{\label{sec:setup}Experimental Setup}

   The experiments were conducted with the rheo-microscope (microscope module) of the Modular Compact Rheometer (MCR-702) from Anton Paar GmbH, Germany, which has a quartz glass-bottom plate of
65 mm diameter. For the upper plate, we used the 25mm diameter steel probe, which is attached to 25mm glass coverslip to ensure the smooth surface. This is done by a PDMS layer deposited using the spin
coater on the glass coverslip by operating at 1000 RPM for 1 minute. This coated coverslip was then carefully placed onto the upper probe such that no air bubbles get trapped between them. This probe was left
at room temperature for 24 hours, followed by heating in the oven for 2 hours at $65^{\circ}$C. This procedure glued the glass coverslip with the rheometer's upper steel plate and made it's surface smooth.
A similar setup was also used for Cell Monolayer Rheology by \citet{Chen:2010jn, Fernandez:2007ch, Dakhil:2016ij} in their studies of bulk rheology of a large collection of single cells, such as fibroblasts and
HeLa cells. In this study, we are using a confluent epithelial cell monolayer with cell-cell contacts.

   The two major challenges faced in the thin gap rheometry are the parallelism of the upper and lower plates of the rheometer and the contact of the upper plate with the material (i.e. cell monolayer in our case).
We managed to overcome both these difficulties (see SI). 
    To ensure the parallelism of the two plates, we first made marks on the bottom plate and then focus on those marks using the microscope with 20X objective. We then fix the focus, which in turn fix the
z-direction. Then, we scan the bottom plate marks from the outer diameter to the centre of the plate (by changing the x and y position) and observe whether any defocussing is happening. We did not observe
any defocussing (see SI). This ensures that the bottom plate is flat and horizontal.  We repeat the same procedure of focussing on the glass coverslip of upper plate and scan for any points/areas
of defocussing. We did not observe any defocusing of the upper plate's glass coverslip (see SI), which ensures that the upper glass coverslip is also flat and parallel to the lower glass plate.

\begin{figure}[htbp]
	\begin{center}
		\includegraphics[width=0.7\columnwidth]{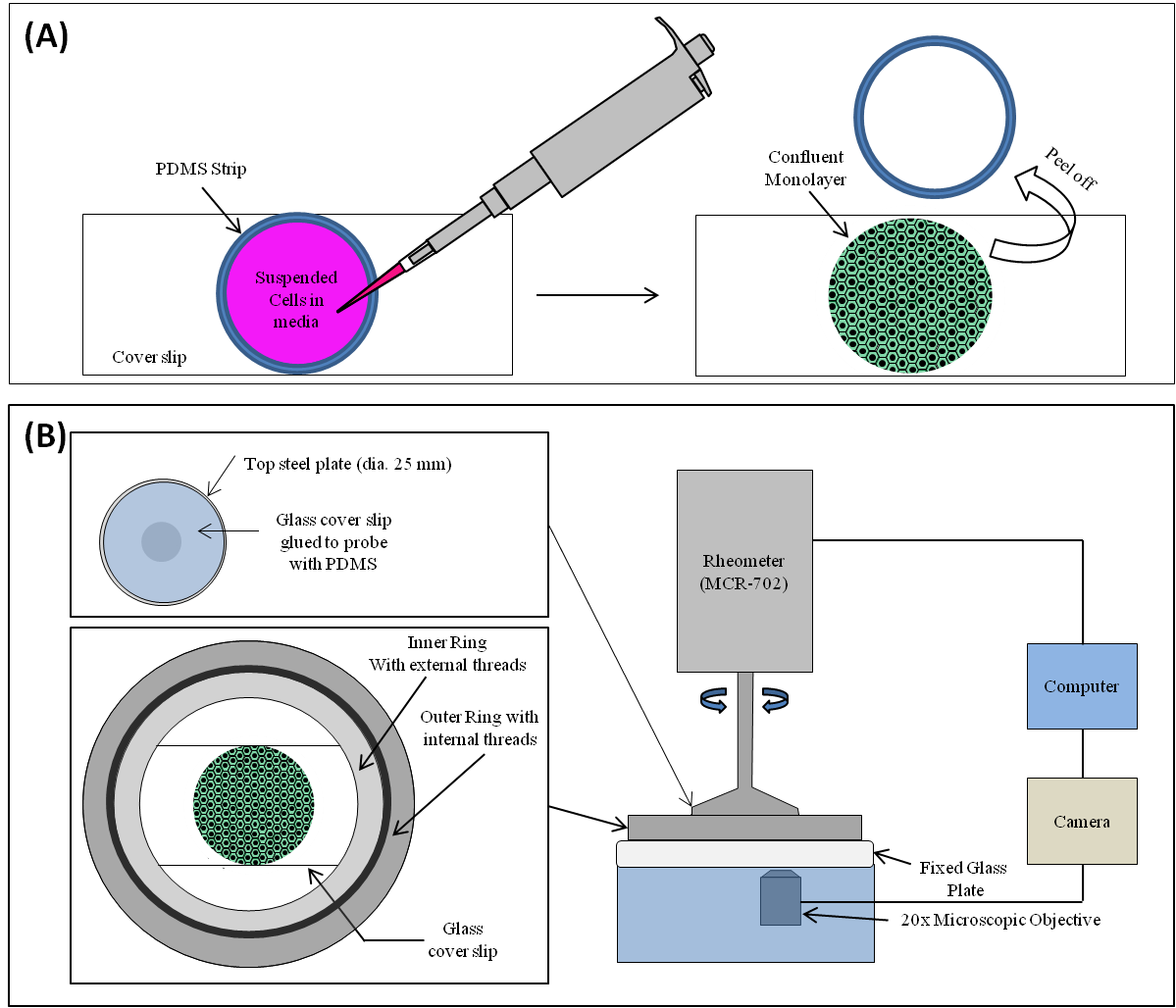}
		\caption{\label{fig:cmr}Schematic representation of the setup to perform rheology of the cell monolayer. (A) preparation of circular cell monolayer using PDMS strip. (B) setup of rheometer for CMR }
	\end{center}
\end{figure}

\begin{figure}[!tbp]
	\begin{subfigure}[b]{0.45\textwidth}
		\includegraphics[width=\textwidth]{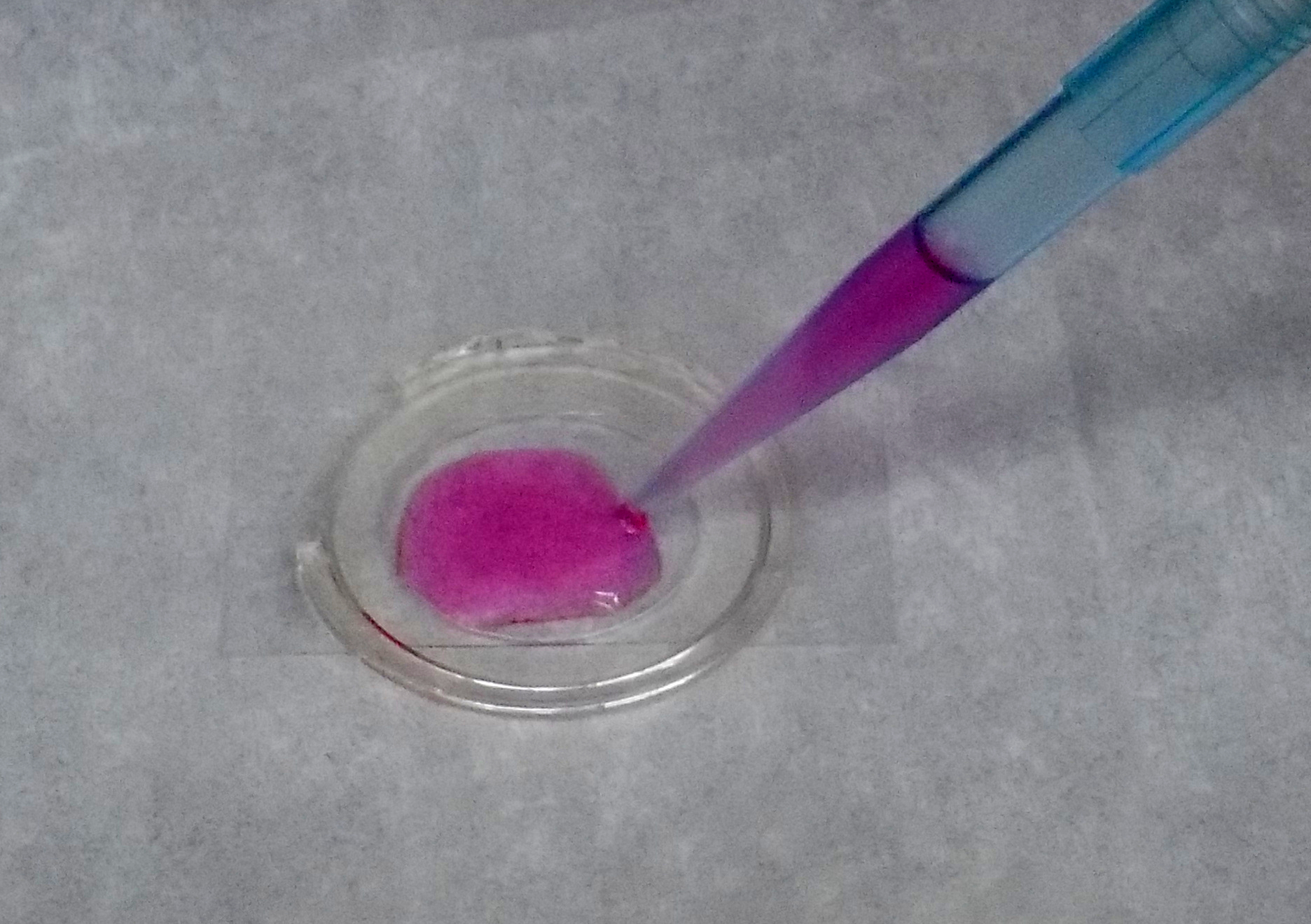}
		\caption{}
		\label{fig:expt1a}
	\end{subfigure}
	\hfill
	\begin{subfigure}[b]{0.45\textwidth}
		\includegraphics[width=\textwidth]{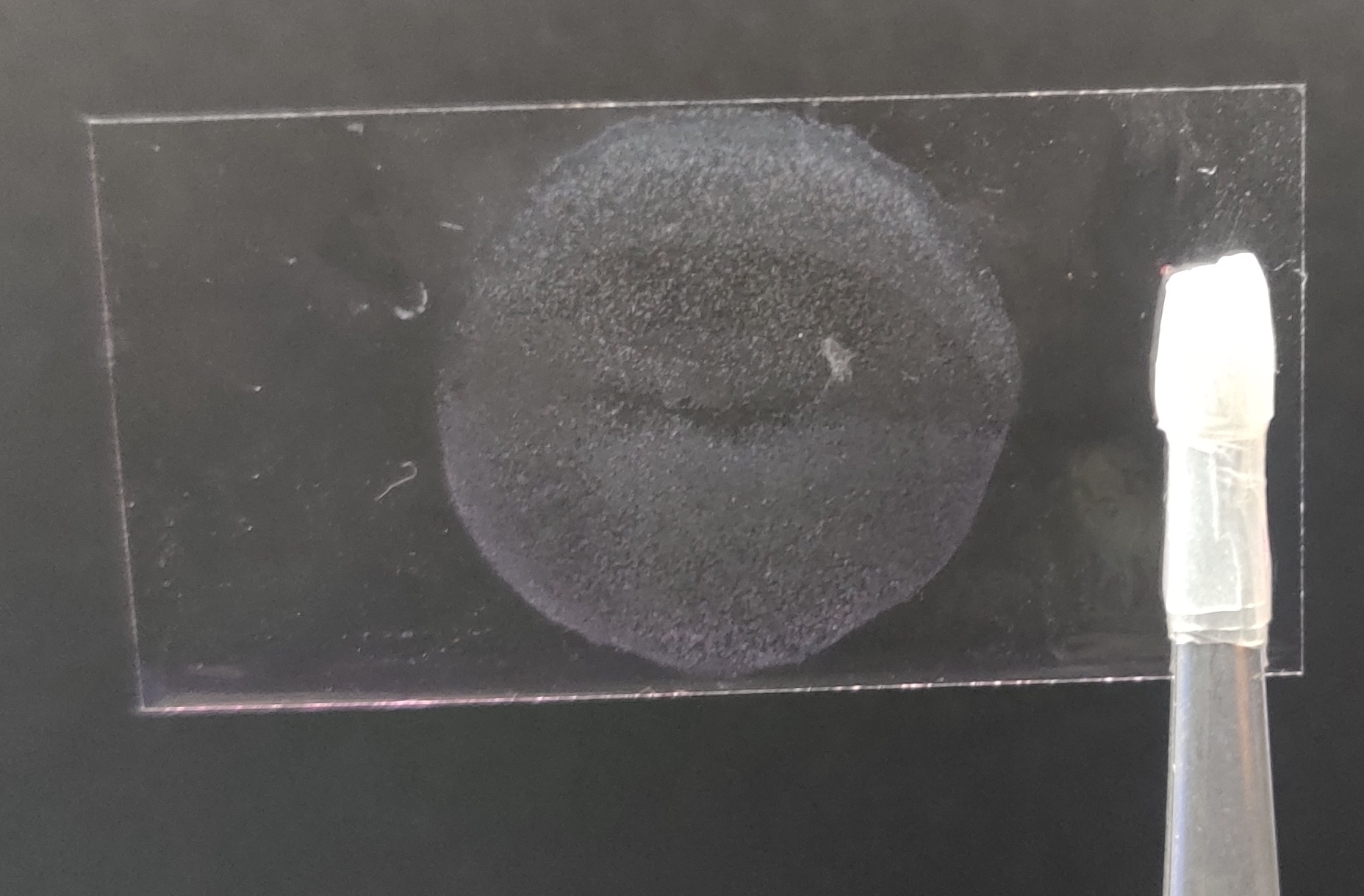}
		\caption{}
		\label{fig:expt1b}
	\end{subfigure}
	\hfill
	\begin{subfigure}[c]{0.45\textwidth}
		\includegraphics[width=\textwidth]{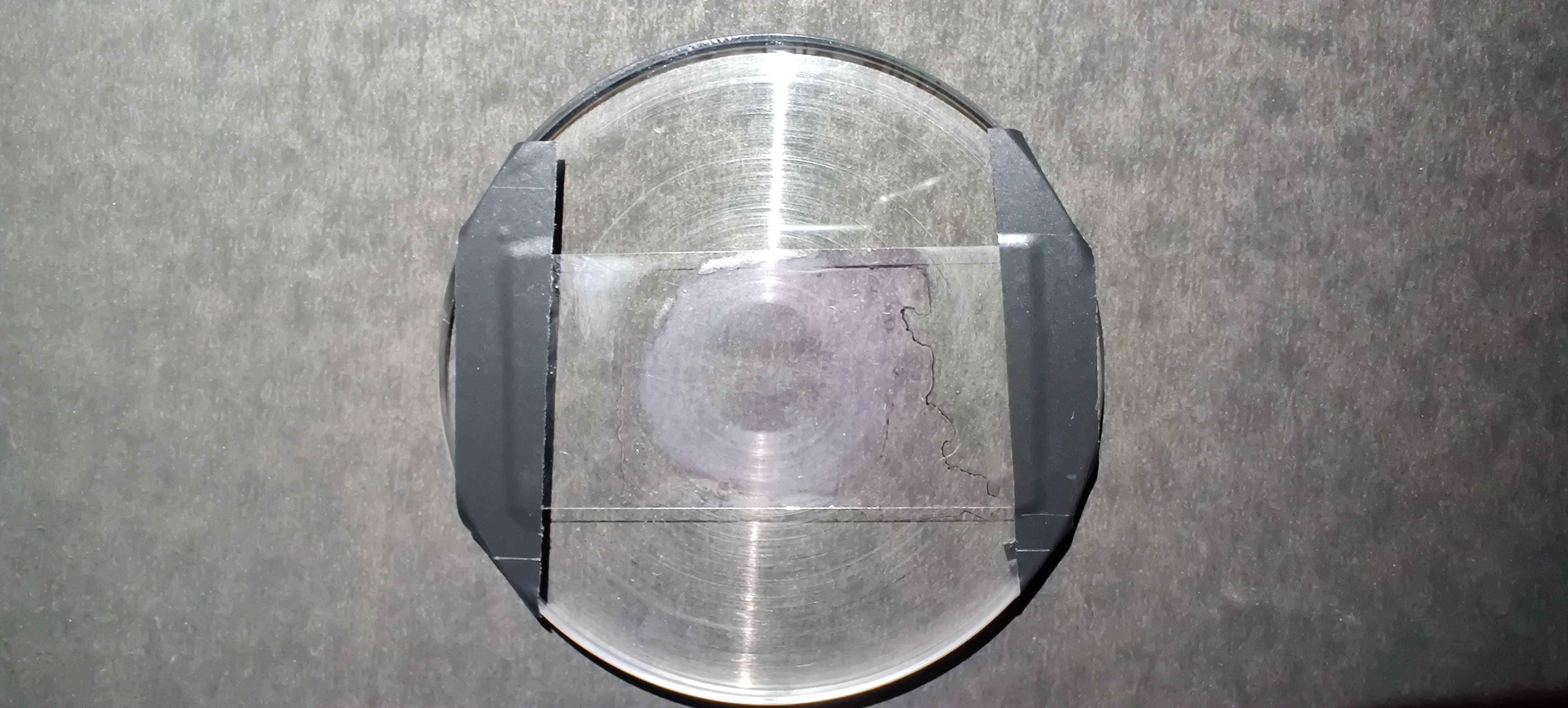}
		\caption{}
		\label{fig:expt1c}
	\end{subfigure}
	\hfill
	\begin{subfigure}[c]{0.45\textwidth}
		\includegraphics[width=\textwidth]{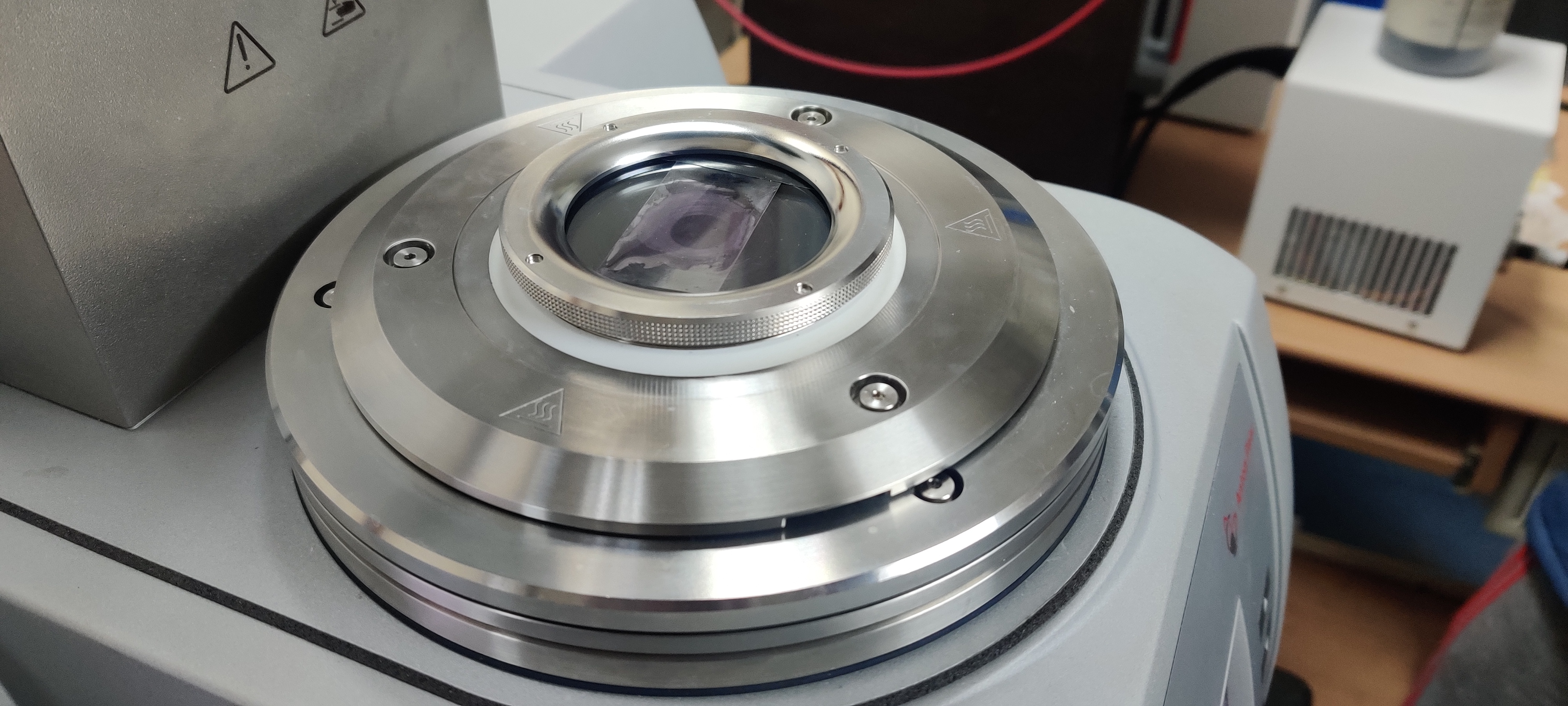}
		\caption{}
		\label{fig:expt1d}
	\end{subfigure}
	\caption{\label{fig:expt1}Mounting of cell monolayer on bottom plate: (a) cell are seeded on rectangular glass coverslip with PDMS ring, (b) the circular confluent cell monolayer after the PDMS ring is
	removed, (c) the rectangular glass coverslip, with circular cell monolayer, is mounted on the bottom glass plate (d) the glass bottom plate with coverslip is mounted on the microscope module of the rheometer.}
\end{figure}

    We have grown the confluent cell monolayer in a 25 mm dia. circle to fully cover the cell monolayer covered by the upper plate of the rheometer. To grow the circular cell monolayer, we created a circular well
using the PDMS ring of the inner diameter of 25 mm on the rectangular coverslip as shown in Fig.~\ref{fig:cmr}A and Fig.~\ref{fig:expt1a}. The cells were seeded in this well and placed in the incubator
(at $37^{\circ}$C in a 5\% CO$_{2}$). When the fully confluent monolayer is formed, the old media was removed, and the cells were prepared for the experiment by supplying the fresh (0\% FBS or 10\% FBS)
media. We supplied the media with 0\% FBS for the experiments with serum-starved cell monolayer and incubated the cell monolayer at $37^{\circ}$C in a 5\% CO2 in a humidified incubator for 8 hours.
The next step is to remove the PDMS well confining the cell monolayer, and rinse the attached cell monolayer with serum-free media. Thus prepared, the cell monolayer is now ready for the rheology experiments
(Fig.~\ref{fig:expt1b}). For the experiments with healthy cells, we incubated the cell monolayer with full serum (10\% FBS) media for 8 hrs then remove the PDMS well and rinse it with full serum media and take
it under the rheometer for experiments. We remove the PDMS ring when the cell monolayer becomes confluent (Fig.~\ref{fig:expt1b}). The rectangular glass coverslip with cell monolayer is fixed firmly on the
glass bottom plate with the help of strong adhesive tapes which are mounted on the sides as shown in Fig.~\ref{fig:expt1c}. This setup with the rectangular coverslip with circular cell monolayer firmly mounted on
the glass-bottom plate is then mounted on the rheometer's microscope module (Fig.~\ref{fig:expt1d}). 

   The shear rheology was performed by maintaining the initial normal force exerted by the upper plate on the material at $\sim$0.1 N. This ensures that the cell monolayer is being sheared (see SI).

   To check, whether we are actually shearing the cell monolayer, we capture the movies of the cell monolayer during the shearing experiments (see SI). The cell monolayer deformation is distinctly visible,
which confirms that the upper plate is touching the cell monolayer and applying the shear on it.

  Following previous studies of  \citet{Fernandez:2007ch, Dakhil:2016ij}, we conducted all the experiments at 25$^{\circ}$C. The effect of temperature on cell monolayer will be the same for all conditions i.e. with
and without serum. In the current study, we assumed that if the temperature is the same for all conditions, its effect on rheological properties may be the same for all rheological experiments and can be factored out. 
Also, to maintain the cell number density constant, we observe the confluent cell monolayer under a fluorescence microscope and measure the cell number density. When the cell number density reaches approximately
at $\sim$ 2.34 $\times$ 10$^{6}$ cells for a circular cell monolayer of 24 mm diameter, we prepare the coverslip with circular cell monolayer for rheological experiments.

\section{\label{sec:results}Results}
\subsection{\label{sec:saos}Harmonic oscillatory shear}
We start our study of the material properties of cell monolayer by applying sinusoidal strain oscillations. For viscoelastic materials we use
\begin{align}
	\gamma(\omega t) &= \gamma_{0}\sin(\omega t) \\
	\sigma(\omega t) &= G^{\prime}\gamma_{0}\sin(\omega t) + G^{\prime \prime}\gamma_{0}\cos(\omega t) \\
	\tan(\delta) &= \frac{G^{\prime \prime}}{G^{\prime}}
\end{align}

where $\omega$ is the angular frequency of the oscillatory shear $\gamma(\omega t)$, $t$ is the time, $\gamma_{0}$ is the amplitude of the oscillatory shear, $\sigma(\omega t)$ is the 
stress, $G^{\prime}$ is the storage modulus, and $G^{\prime \prime}$ is the loss modulus, and $\delta$ is the loss tangent. We aim to explore the linear and non-linear response of this living system. 

   In all our experiments, the total gap between the rheometer plates is the sum of the thickness of glass coverslip and the cell monolayer, which is $\sim$200$\mu$m. The bottom glass plate is used to
set the zero for the rheometer before starting the experiments. After that, we fix the glass coverslip with circular cell monolayer (Fig.~\ref{fig:expt1b}) on the bottom glass plate as shown in
Fig.~\ref{fig:expt1c} and Fig.~\ref{fig:expt1d}. The actual shear strain on the cell monolayer will now be estimated by scaling the shear strain readings from the instrument, $\gamma \%$, by
initial $\gamma_{0}$.  We scaled the shear strain $\gamma\%$ as:

\begin{equation}
\hat{\gamma} = \frac{\gamma}{\gamma_{0}}
\end{equation}

\subsubsection{\label{sec:amplitude}Amplitude sweep}

\begin{figure}[!tbp]
	\begin{subfigure}[b]{0.5\textwidth}
		\includegraphics[width=\textwidth]{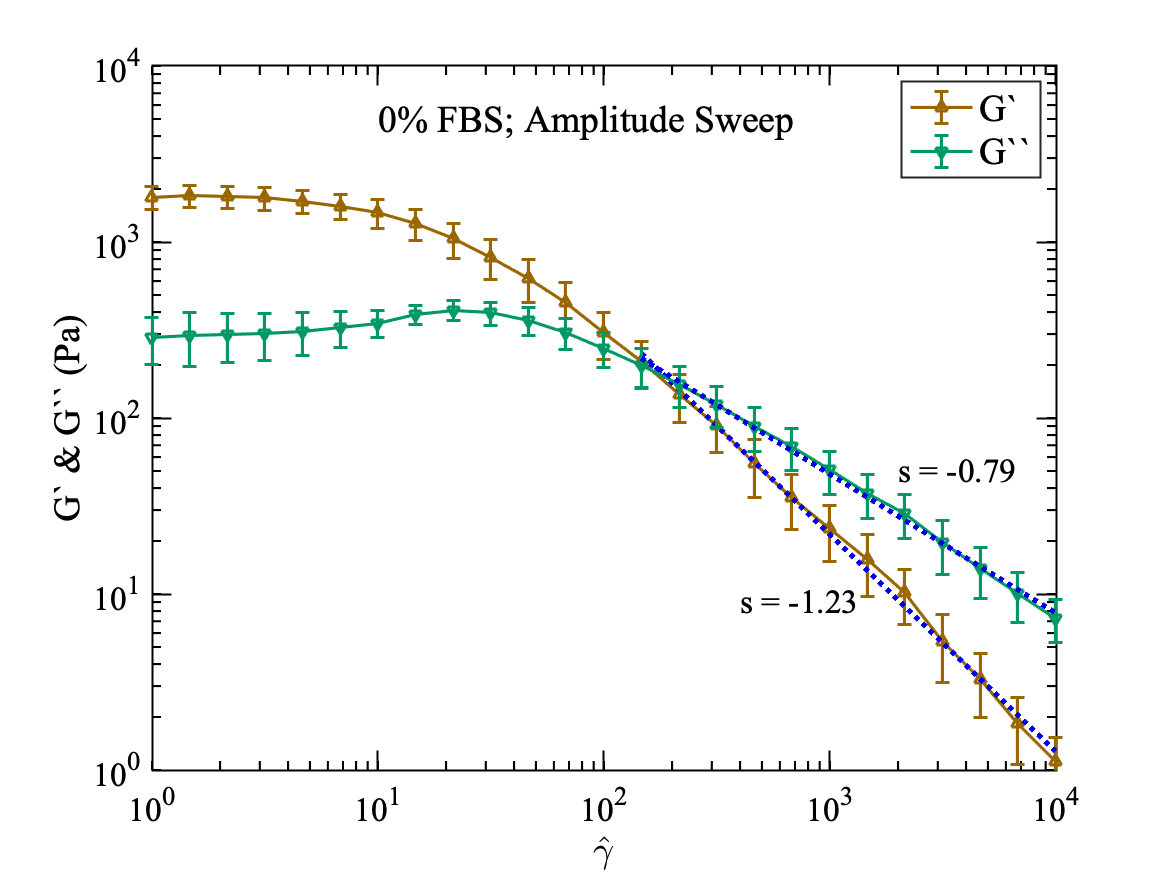}
		\caption{}
		\label{fig:as1}
	\end{subfigure}
	\hfill
	\begin{subfigure}[b]{0.5\textwidth}
		\includegraphics[width=\textwidth]{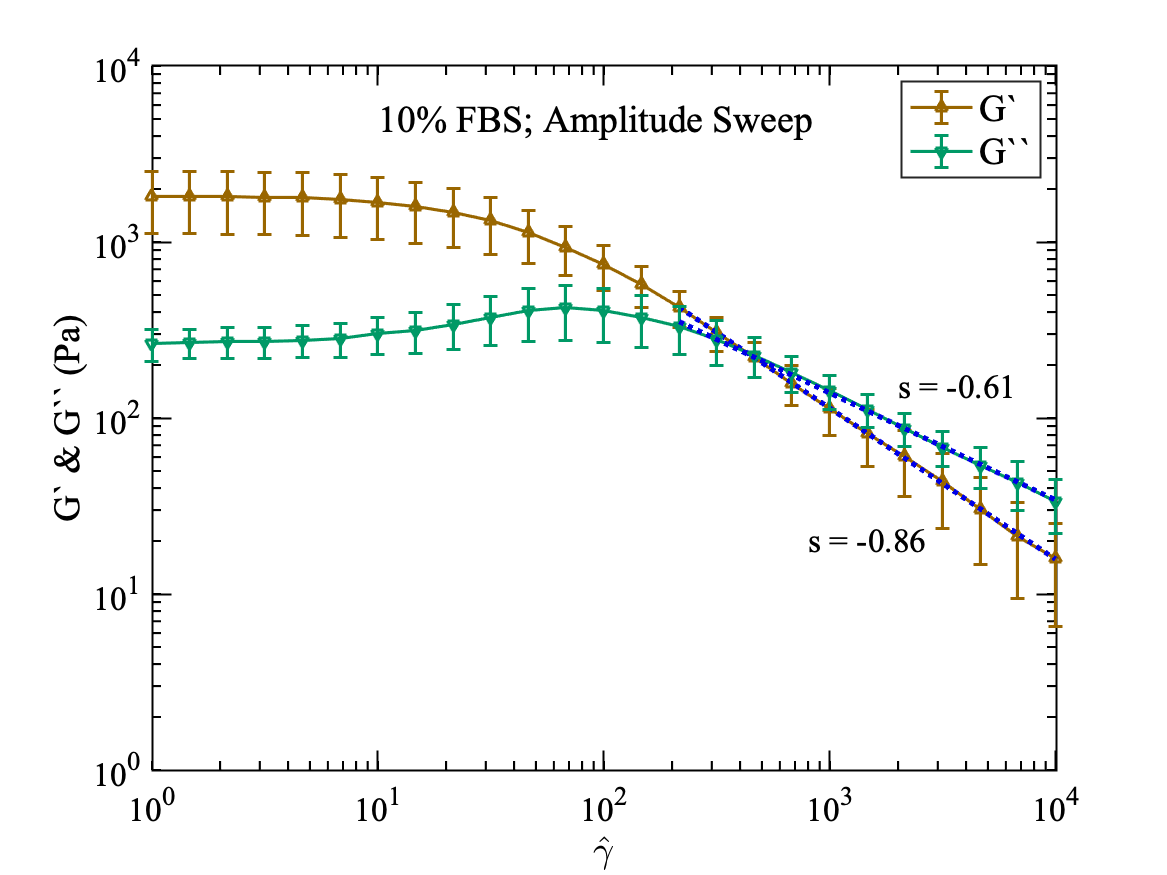}
		\caption{}
		\label{fig:as2}
	\end{subfigure}
	\caption{\label{fig:ampS}Amplitude sweep of the cell monolayer at a frequency of 5 rad/s. The shear moduli ($G^{\prime}$  \&
		$G^{\prime \prime}$) of monolayer with (a) 0\% FBS; (b) 10\% FBS concentration as a function of strain ($\hat{\gamma}$).}
\end{figure}

For the amplitude sweep experiments, the shear strain ($\gamma$) was applied from 0.01\% to 100\% which corresponds to scaled shear strain  $\hat{\gamma}$ from 1 to 10,000 at a frequency of 
$\omega$ = 5 rad/s on the cell monolayer for both cases (a) without serum (FBS = 0\%) and (b) with serum (FBS = 10\%). Moreover, in both cases, the value of storage modulus ($G^{\prime}$) and loss
modulus ($G^{\prime \prime}$) remains constant up to $\hat{\gamma} \sim 8\%$. This is known as the linear viscoelastic (LV) range, and then both moduli keep on decreasing with the increase of the shear
strain (Fig.~\ref{fig:ampS}). The values of $G^{\prime}$ and $G^{\prime \prime}$ are slightly higher for cell monolayer with serum (FBS = 10\%) than that of without serum (FBS = 0\%). This may be
because, under nutrient-deficient conditions, the biopolymers in the cytoskeleton of cells lost their ability to restructure in response to external deformation or strain. Also, the cell-cell contacts in cell monolayer
may be weaker than the healthy cells due to lack of serum. 

    After the linear viscoelastic (LV) range, the storage modulus $G^{\prime }$ starts decreasing. The $G^{\prime}$ decays with a higher rate as compared to $G^{\prime \prime}$, and a tipping
or crossover point occurs after which $G^{\prime \prime}$ become higher than $G^{\prime}$. This crossover occurs at $\hat{\gamma} \sim 210\%$ for starving cell monolayer i.e. at FBS = 0\% 
(Fig.~\ref{fig:as1}) and $\hat{\gamma} \sim 450\%$  for healthy cell monolayer i.e. at FBS = 10\% (Fig.~\ref{fig:as2}). Before the crossover point, the cell monolayer behaves like an elastic solid
(storage modulus dominant over loss modulus). However, after the crossover point, it becomes like a viscous fluid (loss modulus dominant over storage modulus). The transition from elastic solid-like to
viscous fluid-like behaviour is observed at a sufficiently high value of strain. As for small strain values, the cell monolayer may be stretching without significant remodelling of the cell cytoskeleton
(biopolymer network in the cell). However, with the increase of strain values, there may be a significant restructuring of the cytoskeleton in the cells (biopolymer network in the cells start breaking), resulting
in the decrease of both the storage and loss moduli. This can be attributed to the fact that biopolymer linking has distributed strength. First, the links with the smallest strength start breaking, resulting in the
reduction of the rheological properties of the monolayer.  
    
    Furthermore, as shown in Fig.~\ref{fig:ampS}, the storage and loss moduli show power-law dependence on the applied strain after the crossover. The loss modulus, $G^{\prime \prime}$, shows the
power-law exponents of -0.61 and -0.79 for both cases of the cell monolayer with (10\% FBS) and without serum (0\% FBS), respectively. The storage modulus $G^{\prime }$ shows the stronger power-law
dependence with the exponents of -0.86 and -1.23 for cell monolayer with 10\% FBS and 0\% FBS, respectively. The rate of reduction of the value of the storage modulus is more than the loss modulus for both
cell monolayer cases with and without serum (Fig.~\ref{fig:ampS}), as some of the broken contacts of biopolymers may help in maintaining the frictional losses (increases the loss modulus). Hence, the loss
modulus starts dominating over the storage modulus after a particular strain value.  

 As the serum concentration is increased, the dynamic linking of biopolymers increases, resulting in higher strength and restructuring capacity of the monolayer (see Fig.~\ref{fig:ampS}).
Serum starvation reduces the ability of the cell to form new links in the structure of cytoskeleton, and when these links are broken due to deformation of the cells, they are not rebuilding as effectively as a
healthy cell monolayer with 10\% FBS. This results in the early transition of serum-starved cell monolayer from elastic solid-like to viscous fluid-like behaviour.

\begin{figure}[!tbp]
	\begin{subfigure}[b]{0.5\textwidth}
		\includegraphics[width=\textwidth]{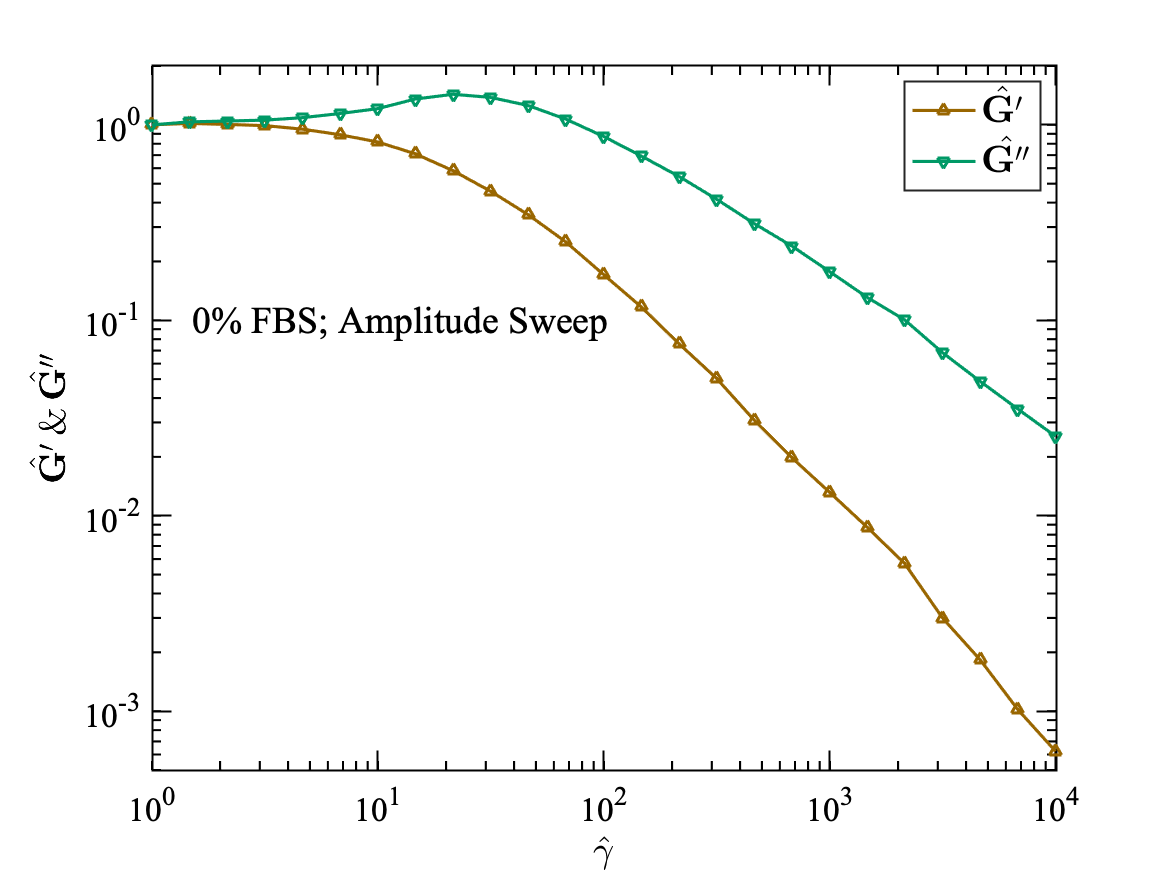}
		\caption{}
		\label{fig:ndas1}
	\end{subfigure}
	\hfill
	\begin{subfigure}[b]{0.5\textwidth}
		\includegraphics[width=\textwidth]{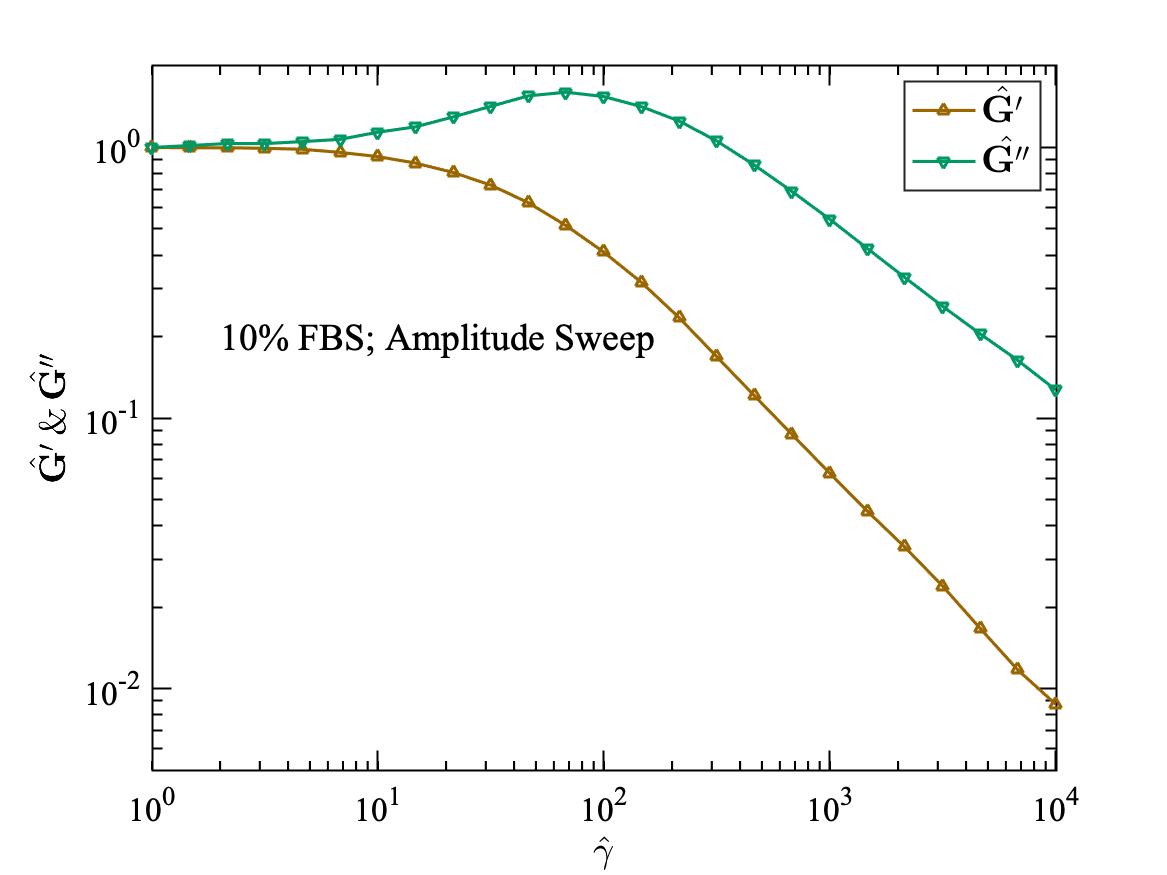}
		\caption{}
		\label{fig:ndas2}
	\end{subfigure}
	\caption{\label{fig:ampSND}Amplitude sweep of the cell monolayer at a frequency of 5 rad/s. The normalized shear moduli ($\hat{\text{G}^{\prime}}$  \&
		$\hat{\text{G}^{\prime \prime}}$) of monolayer with (a) 0\% FBS; (b) 10\% FBS concentration as a function of strain ($\hat{\gamma}$).}
\end{figure}

 Further insight into the material behaviour can be investigated by plotting the normalized storage modulus ($\hat{\text{G}^{\prime}} = G^{\prime}/G^{\prime}(0)$) and the
normalized loss modulus ($\hat{\text{G}^{\prime\prime}} = G^{\prime\prime}/G^{\prime\prime}(0)$)  as a function of applied strain \cite{Hyun:2002gz, Hyun:2011kd}. As seen in Fig.~\ref{fig:ampSND},
the cell monolayer shows a pronounced local maximum in the loss modulus ($\hat{\text{G}^{\prime\prime}}$). This type of behaviour is called as \emph{weak strain overshoot} and is the signature behaviour of soft
glassy materials \cite{Hyun:2002gz, Hyun:2011kd}. \citet{Hyun:2011kd} and \citet{Sim:2003ev} explained the possible mechanism of this behaviour by using a network model composed of segments and junctions. A
segment is defined as a part of a macromolecular chain or a microstructure joining two successive junctions. The junctions are defined as the points where the intra- or intermolecular interactions are localized and
may be regarded as the crosslinking points. The segments are created and lost during the flow and the distribution of junctions is given by their creations and loss rates. 
       The weak strain overshoot behaviour is the result of positive creation and loss parameters, with creation parameter smaller than the loss parameter. With this range of parameters, the creation and the loss terms
increases with the strain amplitude and the destruction rate grows faster than that of creation. The positive creation parameter accounts for the increased connectivity of the network (or other microstructure arising
from interactions) and leads to the increased dissipation, while the loss term becomes dominant at higher strains, which results in overall decrease of both storage ($\hat{\text{G}^{\prime}}$) and
loss ($\hat{\text{G}^{\prime\prime}}$) moduli. The local maximum of the loss modulus $\hat{\text{G}^{\prime\prime}}$ may be the result of the balance between the formation and the destruction of the network
junctions.

   The strain overshoot behaviour in the loss modulus $\hat{\text{G}^{\prime\prime}}$ is highly depends on the class of soft material. For the cell monolayer, the increase of  $\hat{\text{G}^{\prime\prime}}$
is partly related to the destruction of the microstructures developed during the imposed oscillatory shear strain, and the overshoot behaviour is caused by the reformation process of the microstructures.
The local maximum in loss modulus is more pronounced for the healthy cell monolayer with 10\% FBS (Fig.~\ref{fig:ndas2}) than for the starved cell monolayer (Fig.~\ref{fig:ndas1}), again confirming that in
the absence of the serum, the cell monolayer looses the ability to restructure itself.

\begin{figure}[htbp]
	\begin{center}
		\includegraphics[width=0.5\textwidth]{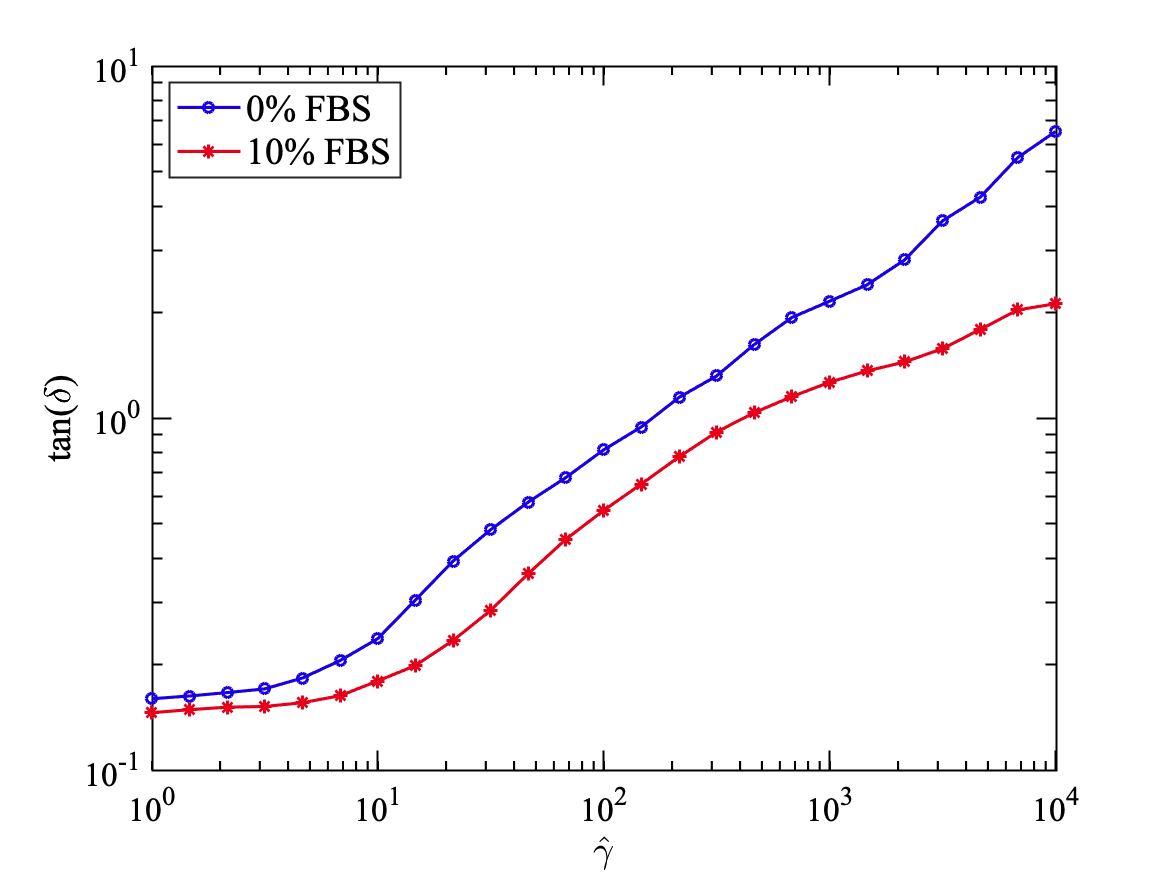}
		\caption{\label{fig:tanD}The value of $\tan(\delta)$ = $G^{\prime \prime}$/$G^{\prime}$ versus strain ($\gamma$ \%), at 0\% and 10\% FBS concentration.}
	\end{center}
\end{figure}

This can also be seen in the graph of $\tan(\delta) = G^{\prime \prime}/G^{\prime}$ in Fig.~\ref{fig:tanD}. After the crossover point, the values of $\tan(\delta)$ become more than unity and keep on
increasing afterwards (Fig.~\ref{fig:tanD}). The values of   $\tan(\delta)$ for serum-starved (0\% FBS) cell monolayer are always higher than that of healthy (10\% FBS) cell monolayer throughout the
range of applied strain. Healthy cells have dynamic cytoskeleton with many different kinds of protein fibres such as actin, myosin, microtubules, intermediate filaments, which gives rigidity and structure to the cells.
In the absence of the serum, cells lack the proteins and other nutrients to form the dynamic filaments, thus reducing the load-bearing capacity and the restructuring capabilities of the cells and the monolayer,
which results in the more fluidic nature of the starved cell monolayer at larger strain values, which results in higher ratios of loss to storage modulus (or $\tan{\delta}$) for starving cell monolayer 
(with 0\% FBS), than that for healthy cell monolayer (with 10\% FBS).

     Many studies such as \citet{Pourati:1998we, Wang:2002bu, Fernandez:2006kt} have shown that cells become stiff with the increasing strain and some others have shown that cells become soft on stretching
\cite{Trepat:2007jb, Krishnan:2009kk}. \citet{Wolff:2012bx} have discussed the paradox of stiffening and softening in cells and reconstituted cytoskeleton gels. Their work has shown that the reconstituted
cytoskeleton network (F-actin/HMM) shows a similar stiffening- softening behaviour. They proposed the inelastic Glassy wormlike chain (iGWLC) model to explain this
peculiar behaviour of cells. They attribute stiffening behaviour to the non-linear stretch of the individual semiflexible biopolymer, which causes the viscoelastic response to the applied stress. The softening is
caused by the dynamical evolution of the mutual bonds between the biopolymers due to the applied strain, characterized as inelastic fluidization. In our work, we believe, that the softening emerges as the
characteristic response to the applied strain because of the evolution of the bonds between the biopolymers and cell-cell contacts in the monolayer.

\subsubsection{\label{sec:frequency}Frequency sweep}

\begin{figure}[!tbp]
	\begin{subfigure}[b]{0.45\textwidth}
		\includegraphics[width=\textwidth]{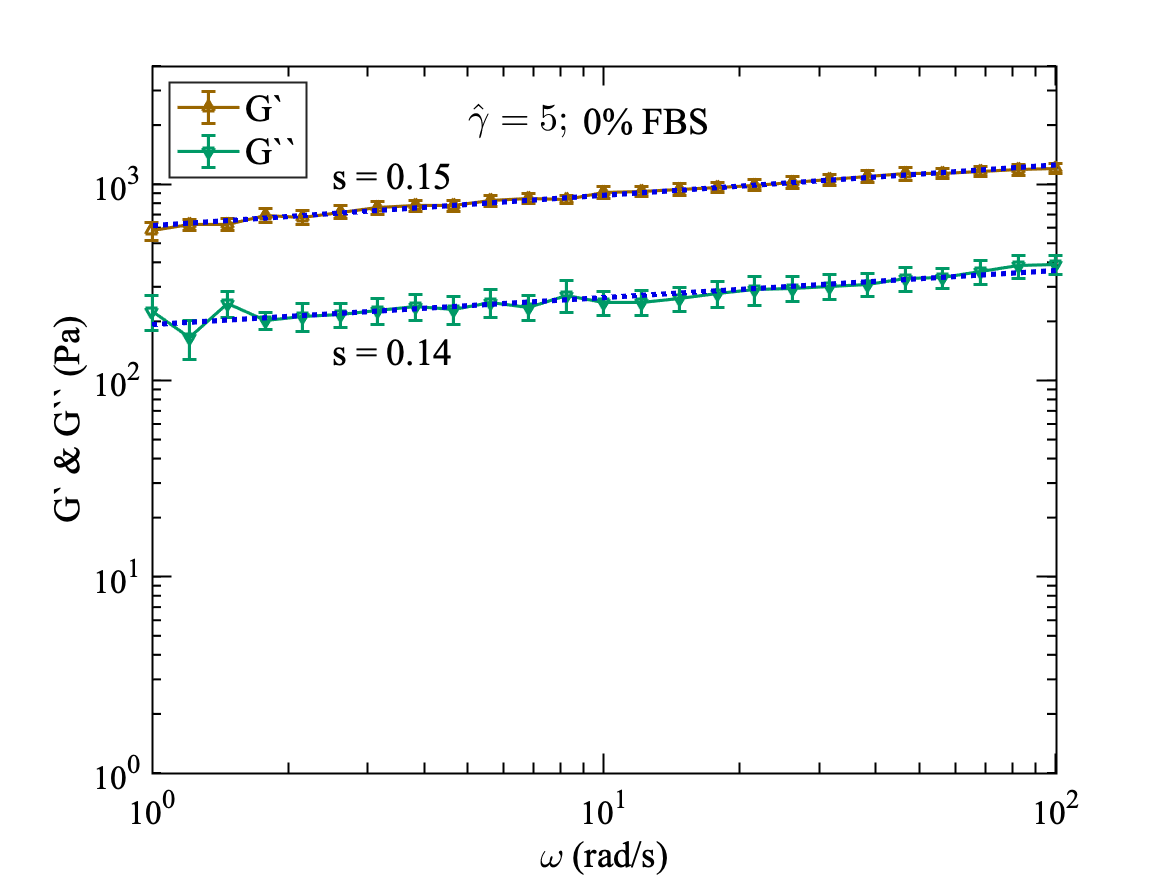}
		\caption{Frequency sweep of the cell monolayer at a constant strain, $\gamma$ = 0.05\%}
		\label{fig:frs1a}
	\end{subfigure}
	\hfill
	\begin{subfigure}[b]{0.45\textwidth}
		\includegraphics[width=\textwidth]{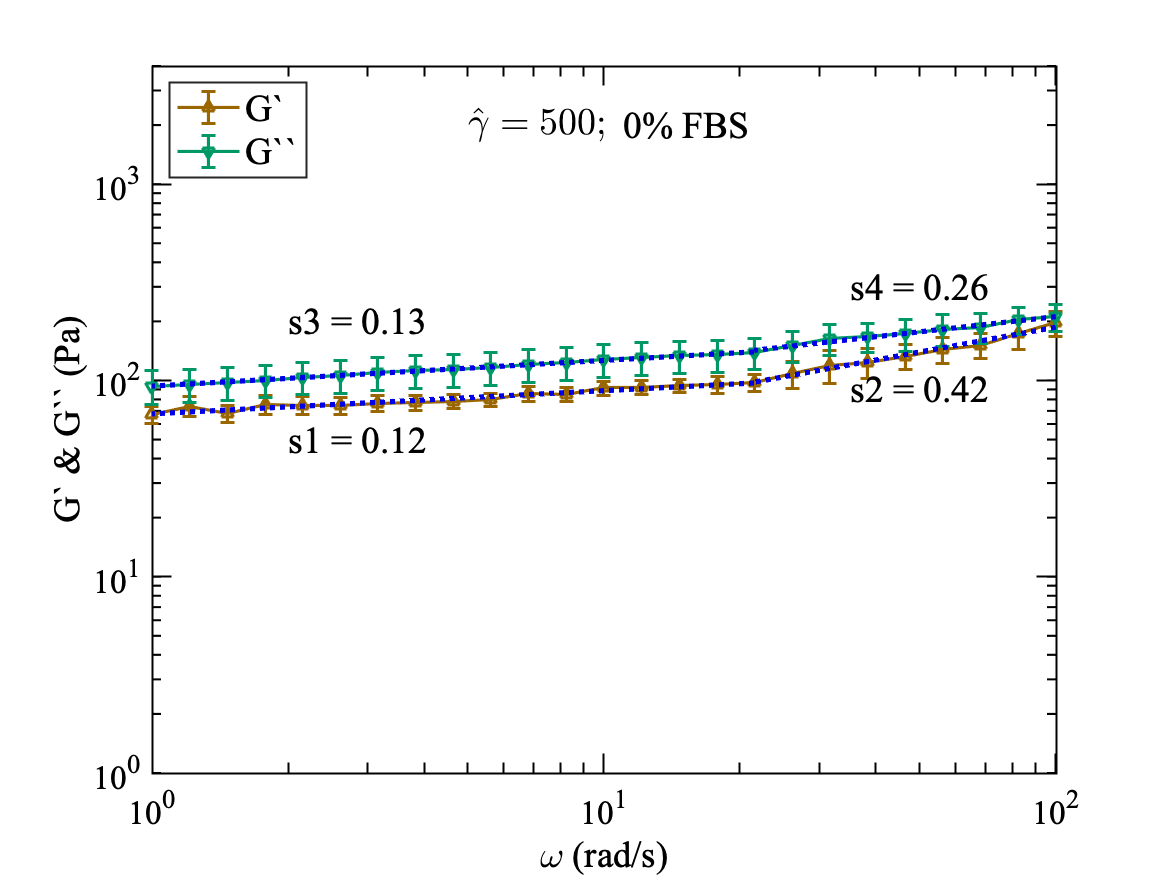}
		\caption{Frequency sweep of the cell monolayer at a constant strain, $\gamma$ = 5\%}
		\label{fig:frs2a}
	\end{subfigure}
	\hfill
	\begin{subfigure}[c]{0.45\textwidth}
		\includegraphics[width=\textwidth]{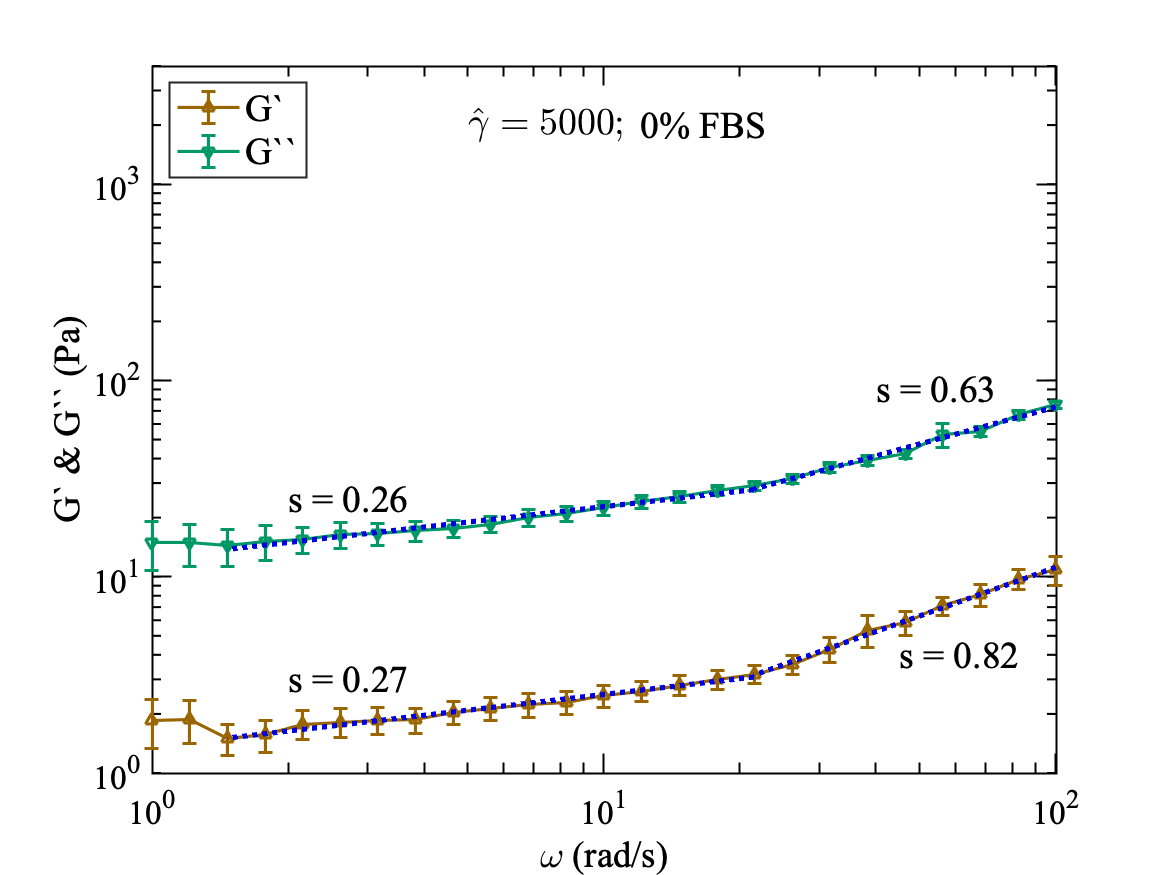}
		\caption{Frequency sweep of the cell monolayer at a constant strain, $\gamma$ = 50\%}
		\label{fig:frs3a}
	\end{subfigure}
	\hfill
	\begin{subfigure}[c]{0.45\textwidth}
		\includegraphics[width=\textwidth]{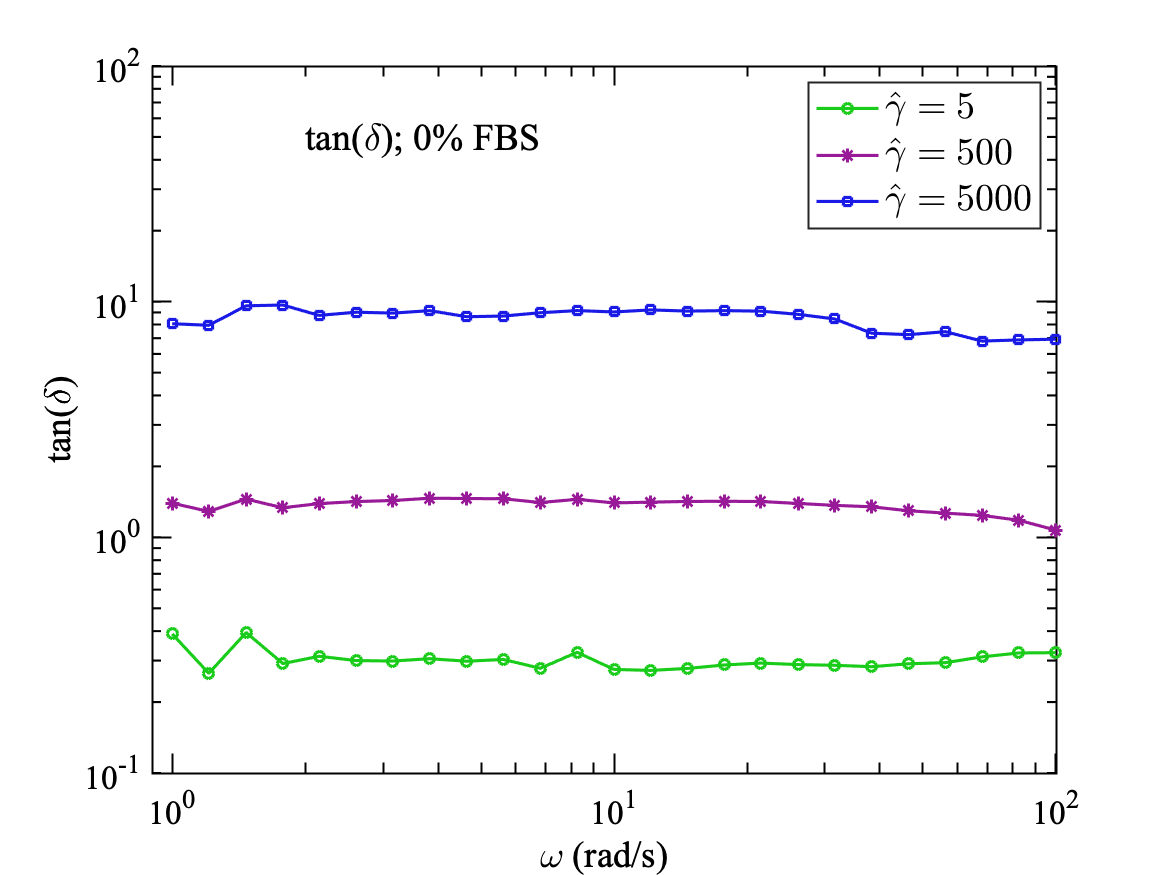}
		\caption{Frequency sweep of the cell monolayer at a constant strain, $\tan{\delta}$}
		\label{fig:frs4a}
	\end{subfigure}
	\caption{\label{fig:frS0}Frequency sweep of the cell monolayer at a constant strain, (a) $\gamma$ = 0.05\% (b) $\gamma$ = 5\%, (c) $\gamma$ = 50\%. 
		(d) $\tan{\delta}$. The shear moduli ($G^{\prime}$  \& $G^{\prime \prime}$) of the monolayer with 0\% FBS concentration as a function of frequency, $\omega$ (rad/s).}
\end{figure}

\begin{figure}[!tbp]
	\begin{subfigure}[b]{0.45\textwidth}
		\includegraphics[width=\textwidth]{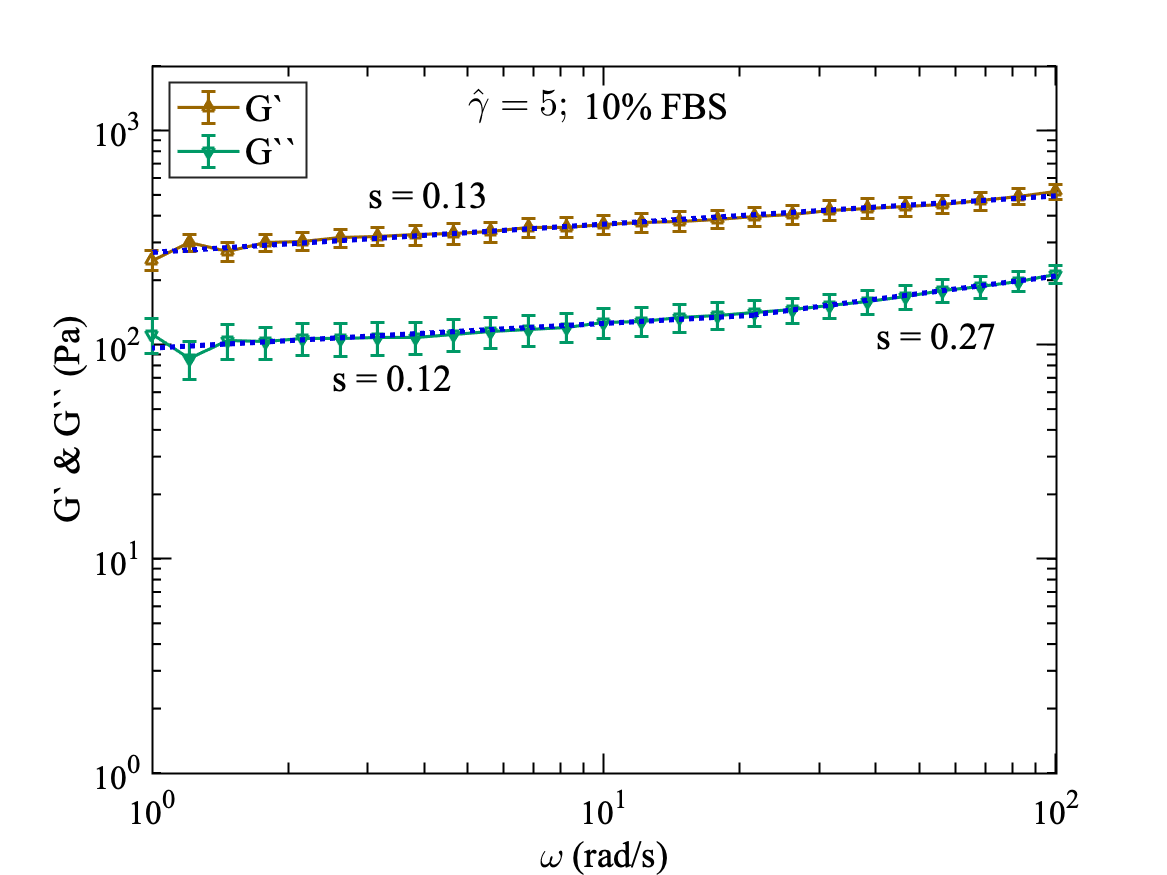}
		\caption{Frequency sweep of the cell monolayer at a constant strain, $\gamma$ = 0.05\%}
		\label{fig:frs1b}
	\end{subfigure}
	\hfill
	\begin{subfigure}[b]{0.45\textwidth}
		\includegraphics[width=\textwidth]{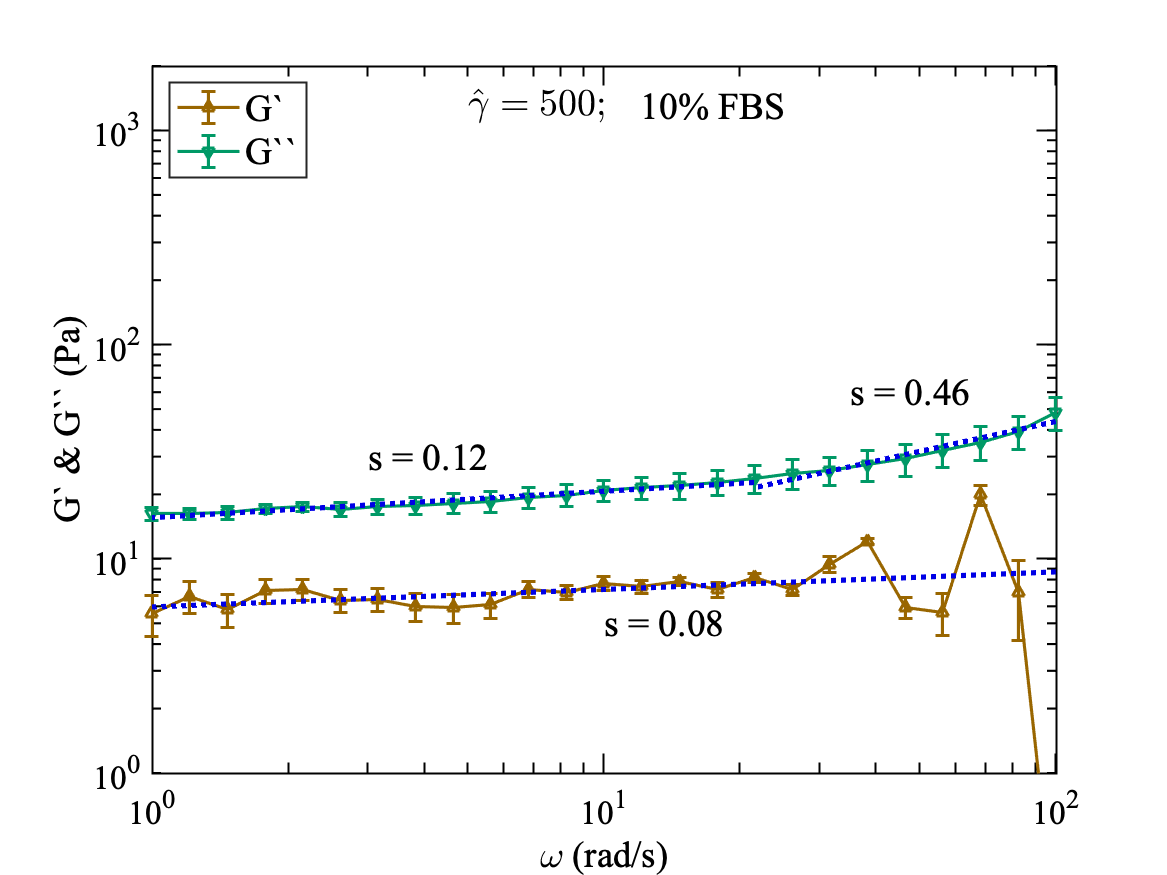}
		\caption{Frequency sweep of the cell monolayer at a constant strain, $\gamma$ = 5\%}
		\label{fig:frs2b}
	\end{subfigure}
	\hfill
	\begin{subfigure}[c]{0.45\textwidth}
		\includegraphics[width=\textwidth]{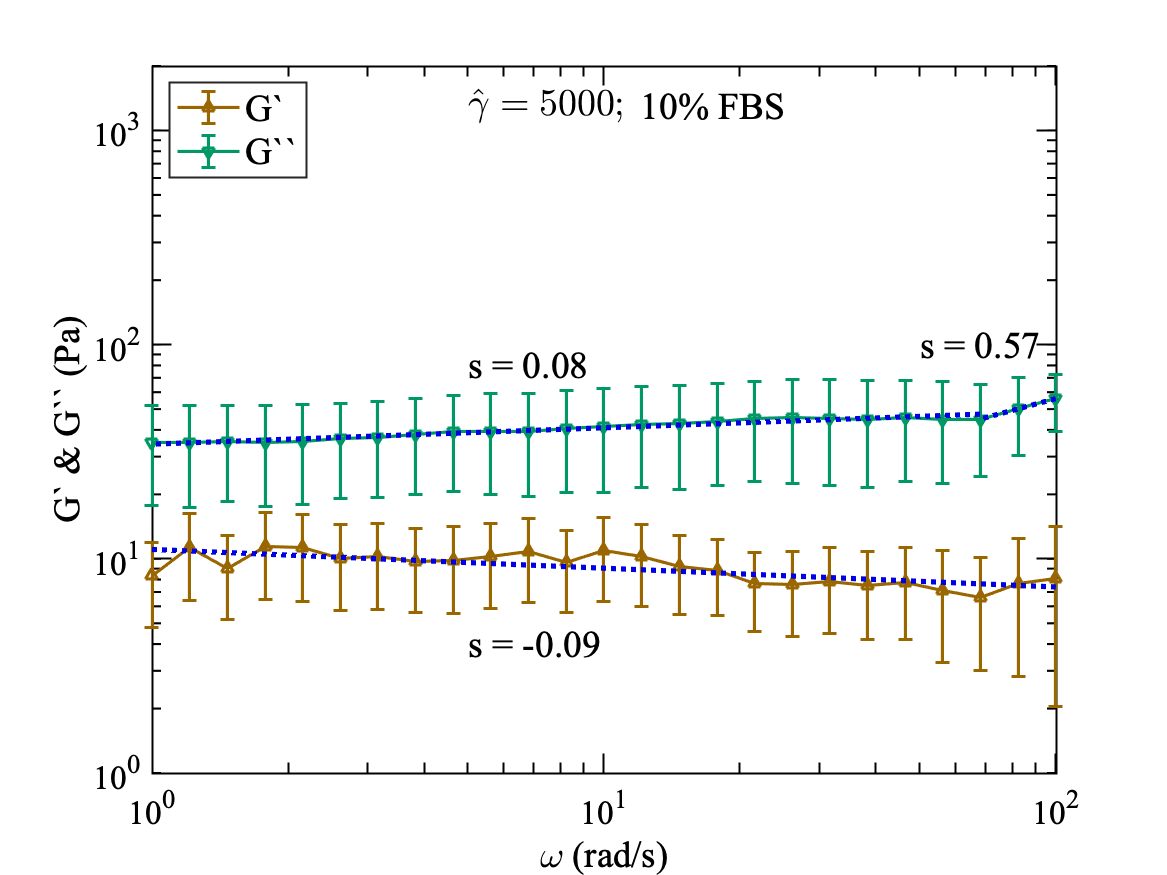}
		\caption{Frequency sweep of the cell monolayer at a constant strain, $\gamma$ = 50\%}
		\label{fig:frs3b}
	\end{subfigure}
	\hfill
	\begin{subfigure}[c]{0.45\textwidth}
		\includegraphics[width=\textwidth]{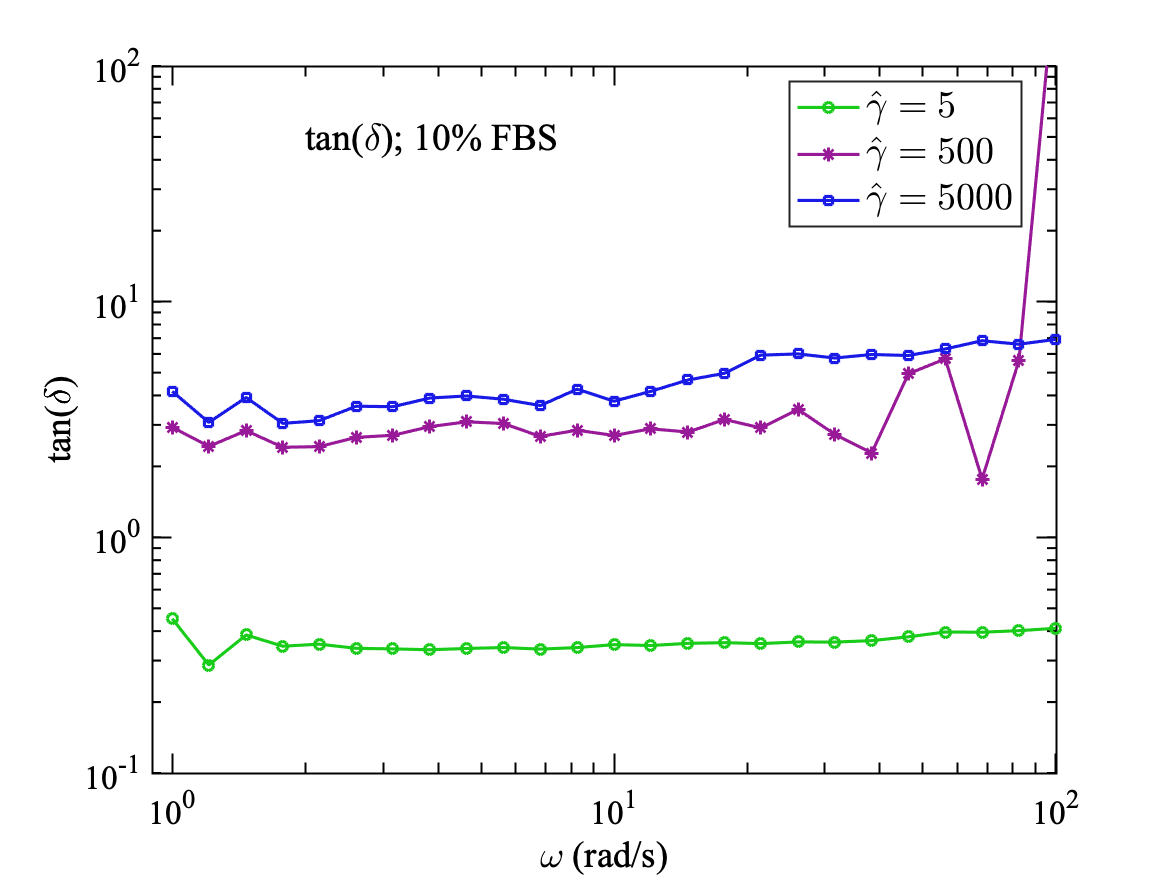}
		\caption{Frequency sweep of the cell monolayer at a constant strain, $\tan{\delta}$}
		\label{fig:frs4b}
	\end{subfigure}
	\caption{\label{fig:frS10}Frequency sweep of the cell monolayer at a constant strain, (a) $\gamma$ = 0.05\% (b) $\gamma$ = 5\%, (c) $\gamma$ = 50\%. 
		(d) $\tan{\delta}$. The shear moduli ($G^{\prime}$  \& $G^{\prime \prime}$) of the monolayer with 10\% FBS concentration as a function of frequency, $\omega$ (rad/s).}
\end{figure}

The frequency sweep experimental data is shown in Fig.~\ref{fig:frS0} for 0\% FBS and Fig.~\ref{fig:frS10} for 10\% FBS. We have applied the frequency sweep in the linear and non-linear viscoelastic 
regime for both cases. We apply the frequency of $\omega$ = 100 to 1 rad/s at a strain amplitude of $\hat{\gamma}$ = 5 in linear regime (Fig.~\ref{fig:frs1a}) and $\hat{\gamma}$ = 500 (Fig.~\ref{fig:frs2a})
and 5000 (Fig.~\ref{fig:frs3a}) in non-linear regime. This procedure is repeated for both cases of the cell monolayer with and without serum.

 The values of both $G^{\prime}$ and $G^{\prime \prime}$ are decreasing as we increase the strain from $\hat{\gamma}$ = 5 to $\hat{\gamma}$ = 500 and $\hat{\gamma}$ = 5000.  In the linear regime at
 $\hat{\gamma}$ = 5, the storage modulus $G^{\prime}$ is higher than the loss modulus $G^{\prime \prime}$ throughout the whole frequency range from $\omega$ = 100 to 1 rad/s. Whereas in the non-linear 
regime at $\hat{\gamma}$ = 500 and $\hat{\gamma}$ = 5000, the $G^{\prime}$ is lower than $G^{\prime \prime}$, which shows the fluid-like behaviour of the cell monolayer in the non-linear regime.
 
 The cell monolayer without serum shows the weak power-law dependence of the storage modulus, $G^{\prime }$ and loss modulus, $G^{\prime \prime}$ with an exponent of $\approx$ 0.15 and $\approx$
0.14 in the linear viscoelastic regime with $\hat{\gamma}$ = 5 throughout the full frequency range (Fig.~\ref{fig:frs1a}), respectively. However, in the non-linear regime, we observe
the stronger dependence of storage and loss moduli at higher frequencies. For $\hat{\gamma}$ = 500, the power-law exponent is 0.12 and 0.13 for $G^{\prime }$  and  $G^{\prime \prime}$ at low frequencies
($\omega \le 20$rad/s), respectively. Both  $G^{\prime }$  and  $G^{\prime \prime}$ show stronger dependence on the frequency at higher frequencies ($\omega \ge 30$rad/s). However,  $G^{\prime }$
shows stronger dependence on the frequency with the power-law exponent of 0.42, while the loss modulus $G^{\prime \prime}$ is having exponent of 0.26 (Fig.~\ref{fig:frs2a}). Fig.~\ref{fig:frs3a} shows
that both moduli at even higher strain ($\hat{\gamma}$ = 5000) show stronger dependence on frequencies than lower strains. At low frequencies below $\omega \leq 20$ rad/s, $G^{\prime }$ and
$G^{\prime \prime}$ are having power-law exponent of 0.27 and 0.26, respectively. At higher frequencies $\omega \geq 30$ rad/s, $G^{\prime }$ shows more dependence than $G^{\prime \prime}$
with power-law exponents of 0.82 and 0.63, respectively. 

Our results are consistent with \citet{Kollmannsberger:2011iy}. They showed that the frequency response of the cells follows a weak power-law over a large range of frequency irrespective of cell type
or experimental technique.  \citet{Miyaoka:2011ej} reported the micro-rheological properties of serum-starved NIH3T3 fibroblasts using AFM. They reported the weak power-law dependence of $G^{\prime }$
on frequency for serum-starved cells, a common characteristic of many other different types of cells. In our study, at the macroscopic level, for serum-starved cell monolayer at 0\% FBS, we
observe the weak power-law behaviour of $G^{\prime } \& G^{\prime \prime}$ at lower frequencies which becomes more pronounced at higher frequencies (Fig.~\ref{fig:frS0}), especially in the non-linear
viscoelastic regime.

  Fig.~\ref{fig:frs4a} shows the variation of $\tan(\delta) = G^{\prime \prime}/G^{\prime}$ with frequency for $\hat{\gamma}$ = 5, 500 and 5000. In the linear viscoelastic regime at $\hat{\gamma}$ = 5,
$\tan(\delta) $ is lowest and does not show dependence with the frequency throughout the applied frequency range $\omega$ = 100 to 1 rad/s. Also, in the non-linear regime,  $\tan(\delta) $ does not depend on
applied frequency, but the values are close to 1 and 10 for $\hat{\gamma}$ = 500 and 5000, respectively. This may be attributed to the fluidic nature of starved cell monolayer at large deformations.

 For full serum case (10\% FBS), the storage modulus $G^{\prime}$ and the loss modulus $G^{\prime \prime}$ show a weak dependence on frequency in the linear and non-linear viscoelastic regime 
throughout the range of applied frequencies. We apply the frequency of $\omega$ = 100 to 1 rad/s at a strain amplitude of $\hat{\gamma}$ = 5 in linear regime (Fig.~\ref{fig:frs1b}) and 
$\hat{\gamma}$ = 500 (Fig.~\ref{fig:frs2b}) and 5000 (Fig.~\ref{fig:frs3b}) in non-linear regime. 

   In linear viscoelastic regime at $\hat{\gamma}$ = 5, (Fig.~\ref{fig:frs1b}) the loss modulus $G^{\prime \prime}$ shows a stronger dependence at high frequencies ($\omega \geq 20$ rad/s) with a power-law
exponent of 0.27. However, at lower frequencies ($\omega \leq 20$ rad/s), $G^{\prime \prime}$ shows weak power-law dependence with an exponent of and 0.12. Whereas, the storage modulus $G^{\prime}$
shows the weaker power-law dependence throughout the applied frequency range with an exponent of 0.13.  In the non-linear viscoelastic regime, at $\hat{\gamma}$ = 500 (Fig.~\ref{fig:frs2b}) the loss
modulus shows similar behaviour in the linear regime, but with lower values. The power-law dependence is weaker at lower frequencies with an exponent of 0.12 and stronger with an exponent of 0.46
at higher frequencies. The storage modulus $G^{\prime}$ also shows the weak dependence with much lower values but with the same power-law exponent of 0.08. The loss modulus is higher than storage modulus, 
indicating the fluidic nature of the material.

  At even higher strain at $\hat{\gamma}$ = 5000 (Fig.~\ref{fig:frs3b}) in the non-linear regime, the storage and loss moduli show very weak dependence on frequencies throughout the applied range. Except for
loss modulus, which shows the power-law dependence with an exponent of 0.57 at very high frequencies. The storage modulus values are slightly higher at low frequencies ($\omega \leq 20$ ) and low values at
$\omega \geq 20$, but the exponent of power-law is -0.09. 
  
  Serum-starved cell monolayer shows a strong power-law dependency at a large strain value ($\hat{\gamma}$ = 5000) and large frequency values (see Fig.~\ref{fig:frS0}). This indicates that the serum-starved
cell monolayer becomes stiff at large values of strain at high frequency. Whereas, the cell monolayer with FBS 10\% remain flexible even at large values of strain and at a large frequency, as power law dependency 
remains weak for the cell monolayer with 10\% FBS. 
  
  The ratio of loss modulus and storage modulus, $\tan{\delta}$ is shown in Fig.~\ref{fig:frs4b} for healthy cell monolayer with 10\% FBS. Similar to the starved cell monolayer at 0\% FBS, the $\tan{\delta}$
in the linear viscoelastic regime at $\hat{\gamma}$ = 5 does not depend on the applied frequency. Similar behaviour can be seen for higher deformation at $\hat{\gamma}$ = 500 and 5000. The values of 
$\tan{\delta}$ for $\hat{\gamma}$ = 5 are lowest compared to $\hat{\gamma}$ = 500 and 5000. However, in case of healthy cell monolayer the values of $\tan{\delta}$ for $\hat{\gamma}$ = 500 and 5000
much closer and approximately close to 4 (Fig.~\ref{fig:frs4b}), whereas, for starved cell monolayer the values $\tan{\delta}$ for $\hat{\gamma}$ = 500 and 5000 are close to 1 and 10 (Fig.~\ref{fig:frs4a}),
respectively. This difference between starved cell monolayer at 0\% FBS and healthy cell monolayer at 10\% FBS is due to the presence of serum which provides the necessary nutrients and proteins to the cells
for restructuring the cytoskeleton inside the cells and the contacts with other cells in the cell monolayer. In the absence of the serum, a cell is unable to form necessary proteins and contacts, thus becoming more
fluidic at large deformations.
  
\subsection{\label{sec:stS}Step Strain Experiments}

  Next, we applied a step strain to the cell monolayer to probe more into the bulk rheology of the cell monolayer. Here we start by applying strain of $\hat{\gamma}$ = 2 for 60 s. and then apply a
step increase in strain at $\hat{\gamma}$ = 8 in the linear viscoelastic regime for next 300 s and then decrease the strain to the initial value of $\hat{\gamma}$ = 2 for next 600 s. For non-linear viscoelastic
regime, we apply $\hat{\gamma}$ = 500  for 300 s. after the initial strain of 2 and then decrease the strain to the initial value of $\hat{\gamma}$ = 2 for next 600 s. This procedure is repeated for
both cases 1) serum-free (0\% FBS) and 2) with serum (10\% FBS).

\begin{figure}[!tbp]
	\begin{subfigure}[b]{0.45\textwidth}
		\includegraphics[width=\textwidth]{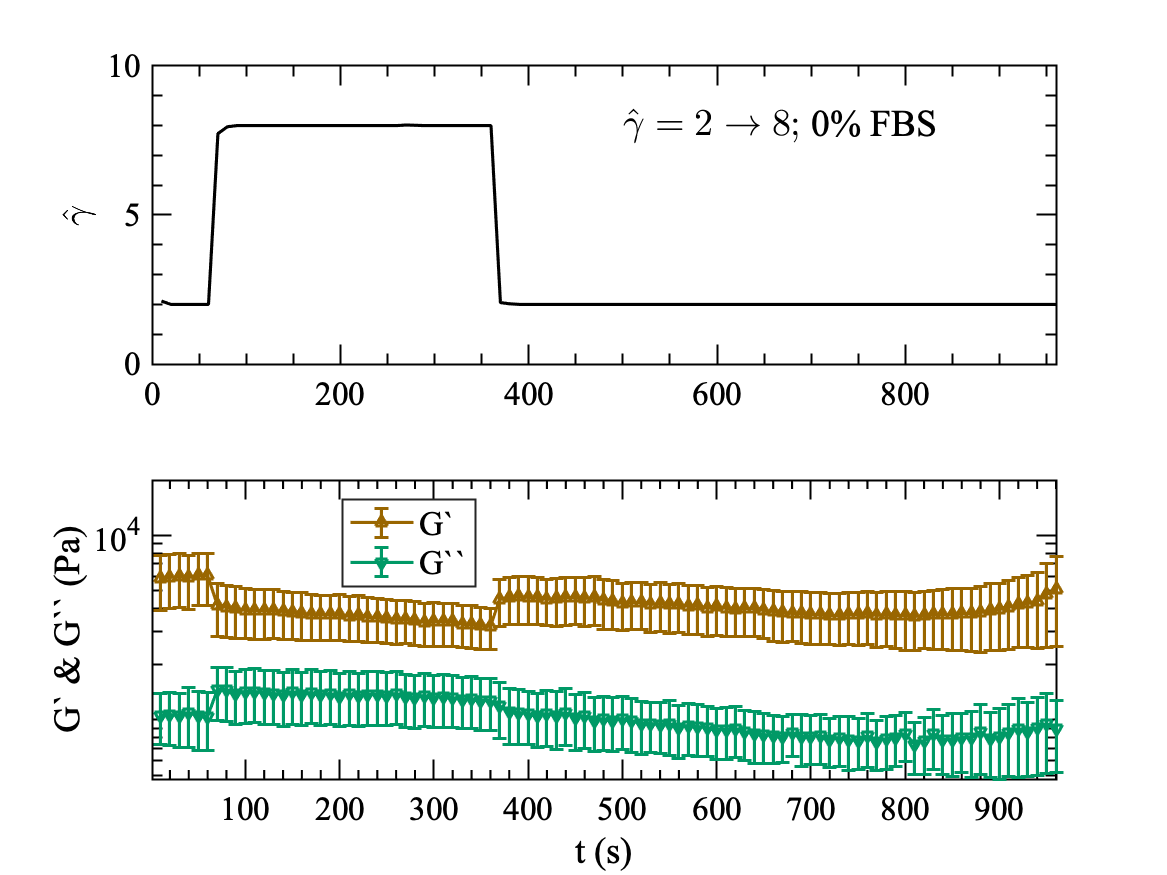}
		\caption{$\hat{\gamma}$ = 2 to 8}
		\label{fig:stS02_08a}
	\end{subfigure}
	\hfill
	\begin{subfigure}[b]{0.45\textwidth}
		\includegraphics[width=\textwidth]{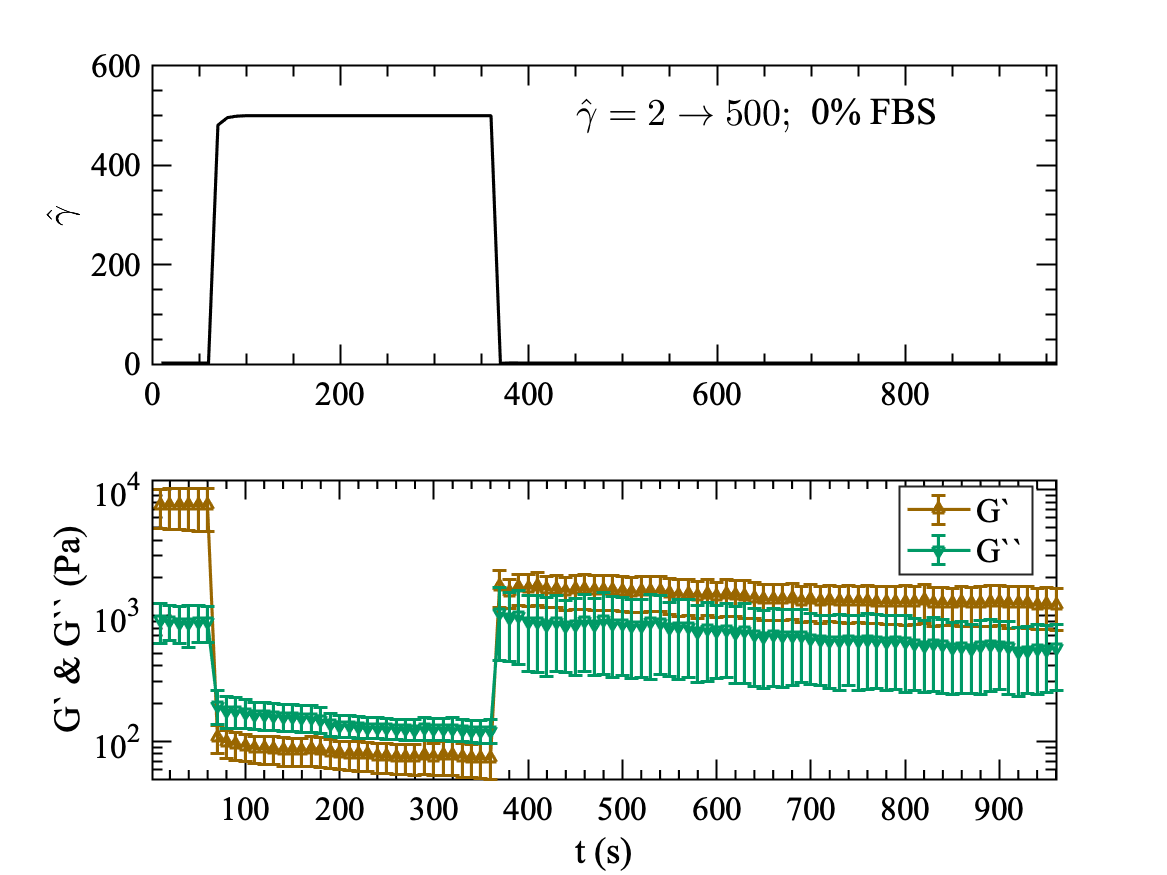}
		\caption{$\hat{\gamma}$ = 2 to 500}
		\label{fig:sts02_5a}
	\end{subfigure}
	\caption{\label{fig:stepStrain1}Step Strain for 0\% FBS}
\end{figure}

\begin{figure}[!tbp]
	\begin{subfigure}[b]{0.45\textwidth}
		\includegraphics[width=\textwidth]{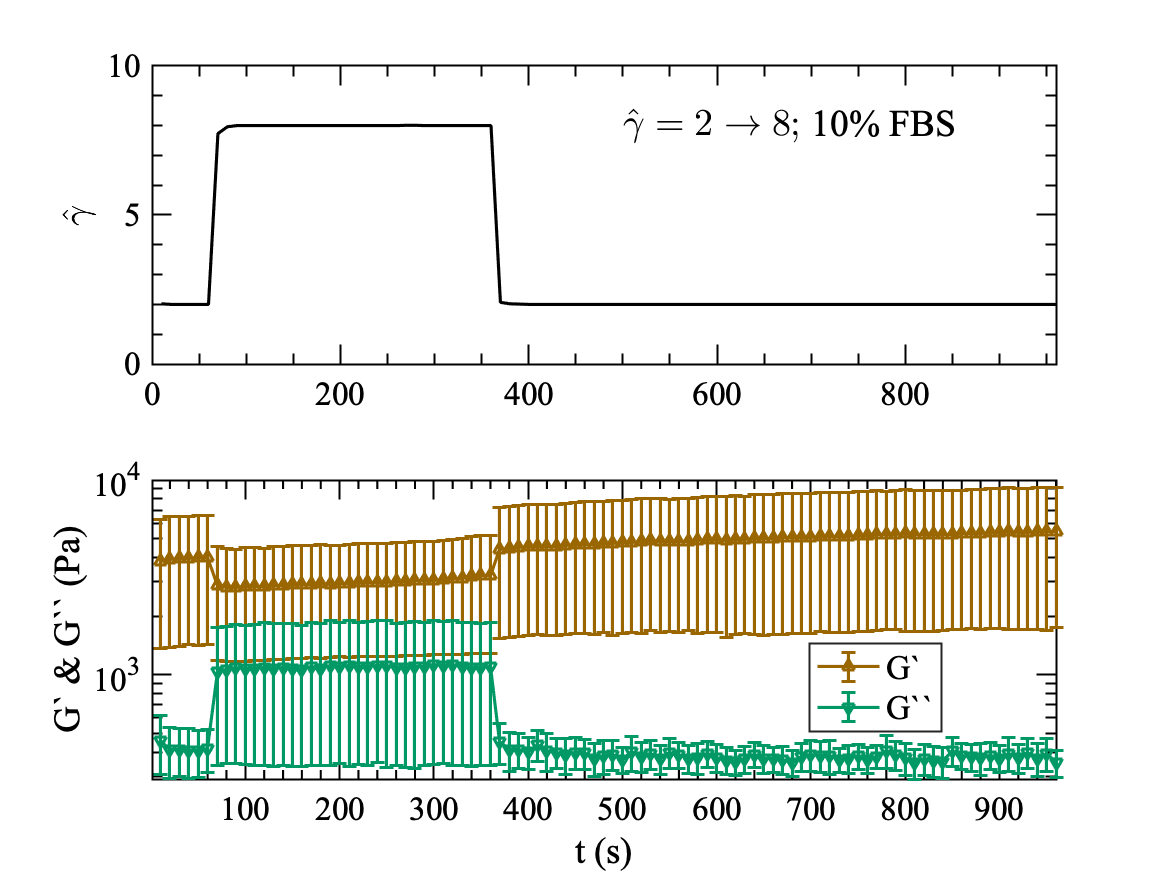}
		\caption{$\hat{\gamma}$ = 2 to 8}
		\label{fig:stS02_08b}
	\end{subfigure}
	\hfill
	\begin{subfigure}[b]{0.45\textwidth}
		\includegraphics[width=\textwidth]{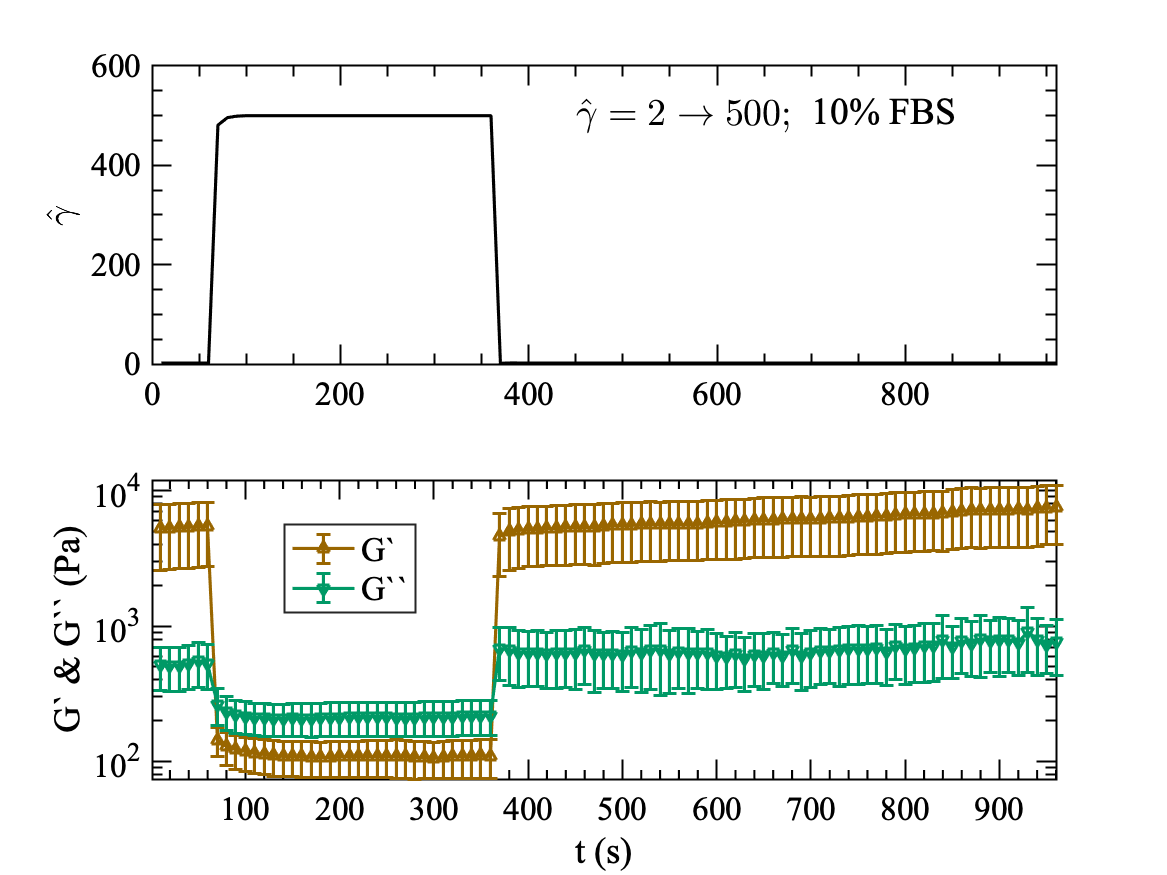}
		\caption{$\hat{\gamma}$ = 2 to 500}
		\label{fig:sts02_5b}
	\end{subfigure}
	\caption{\label{fig:stepStrain2}Step Strain for 10\% FBS}
\end{figure}

\begin{figure}[!tbp]
	\begin{subfigure}[b]{0.45\textwidth}
		\includegraphics[width=\textwidth]{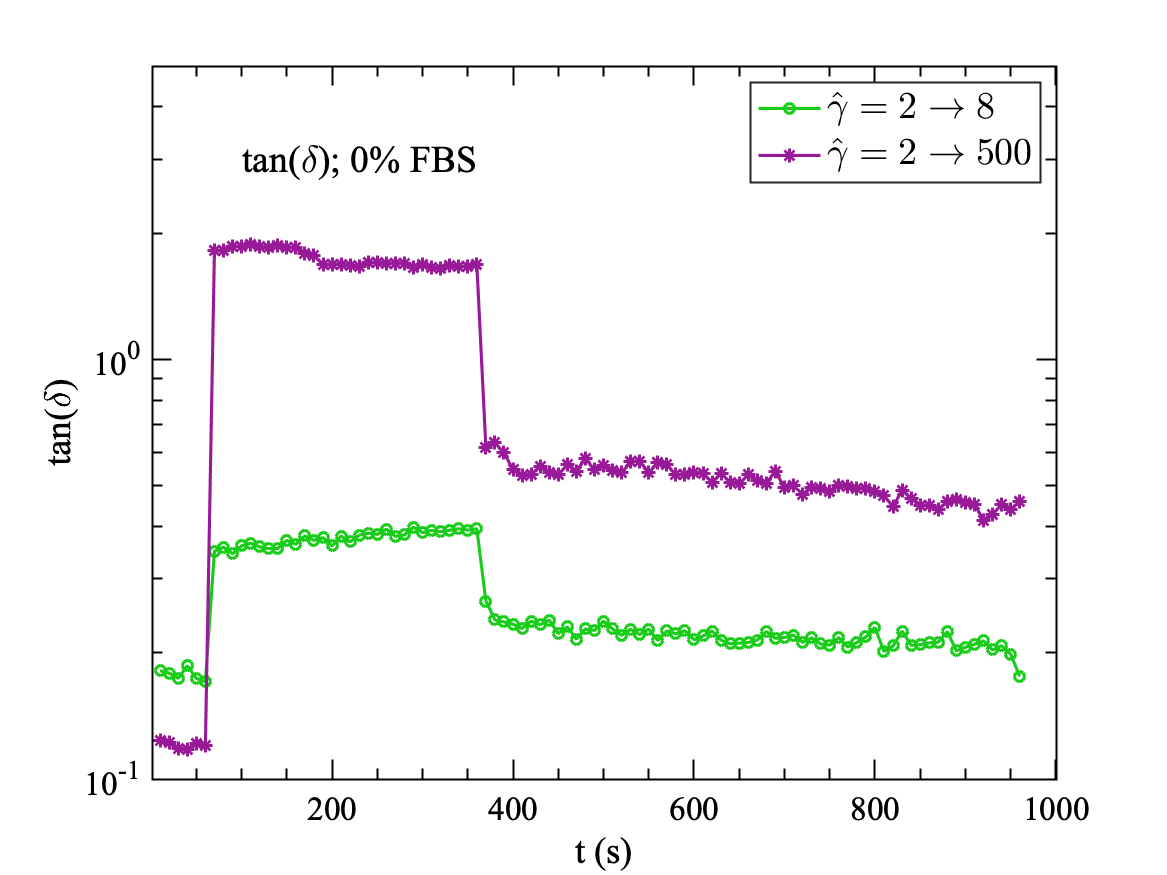}
		\caption{0\% FBS}
		\label{fig:ssTD0fbs}
	\end{subfigure}
	\hfill
	\begin{subfigure}[b]{0.45\textwidth}
		\includegraphics[width=\textwidth]{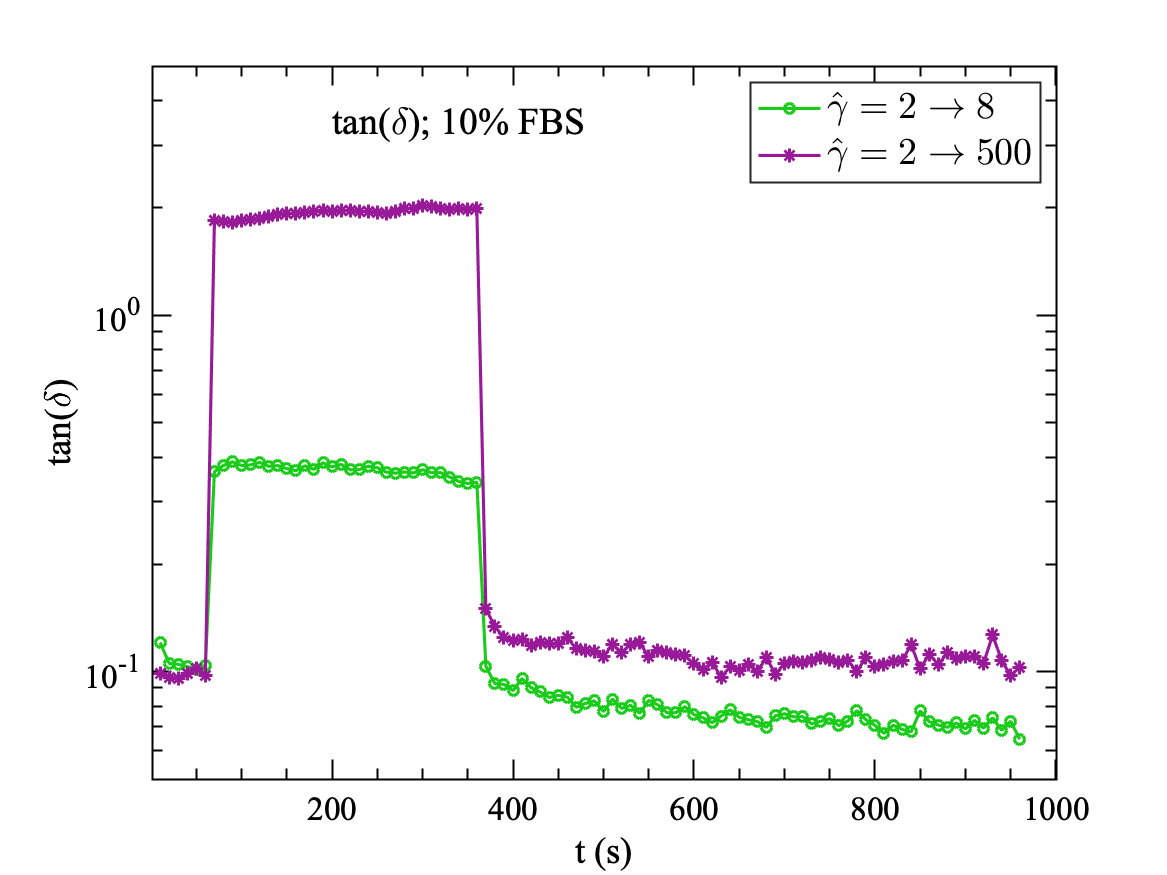}
		\caption{10\% FBS}
		\label{fig:ssTD10fbs}
	\end{subfigure}
	\caption{\label{fig:stepStrain3}Comparison of $\tan{\delta}$ for the linear and non-linear viscoelastic regime}
\end{figure}

\begin{figure}[!tbp]
	\begin{subfigure}[b]{0.5\textwidth}
		\includegraphics[width=\textwidth]{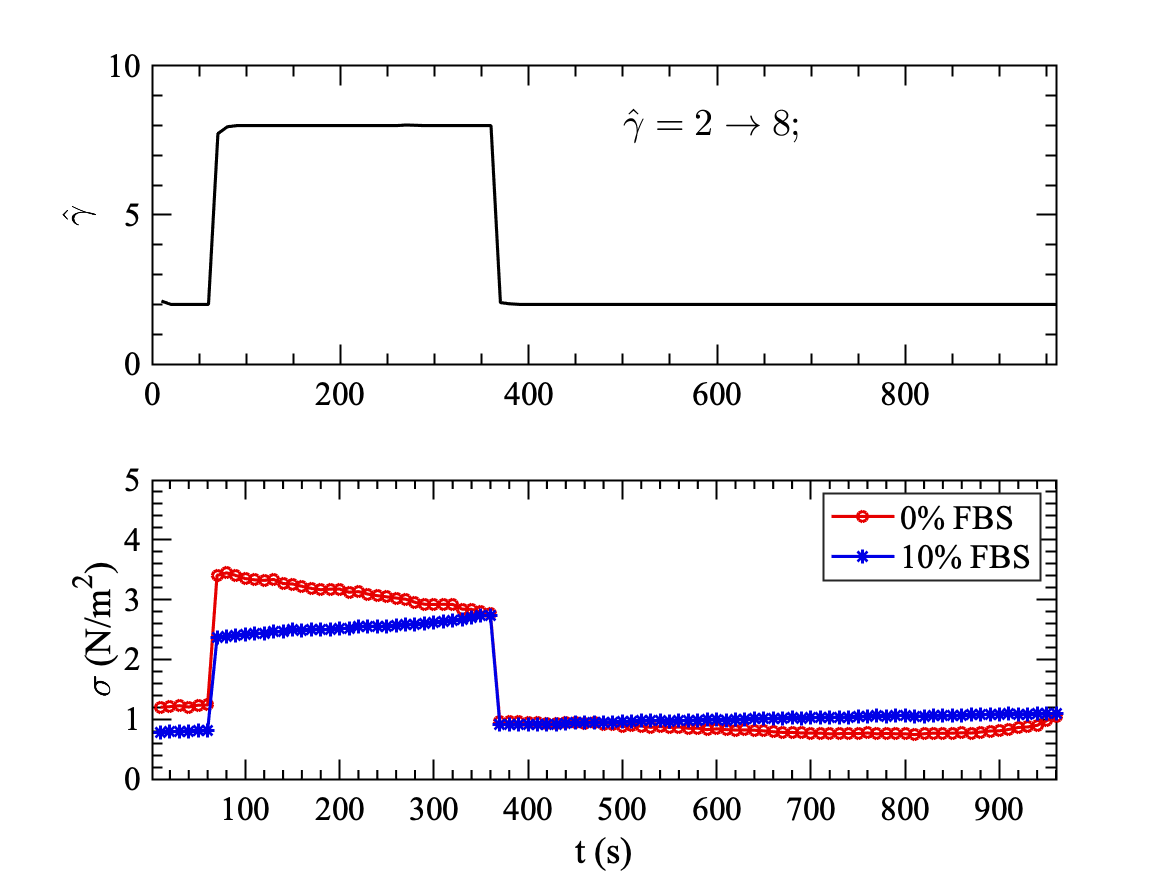}
		\caption{$\hat{\gamma}$ = 2 to 8}
		\label{fig:stress_08ss}
	\end{subfigure}
	\hfill
	\begin{subfigure}[b]{0.5\textwidth}
		\includegraphics[width=\textwidth]{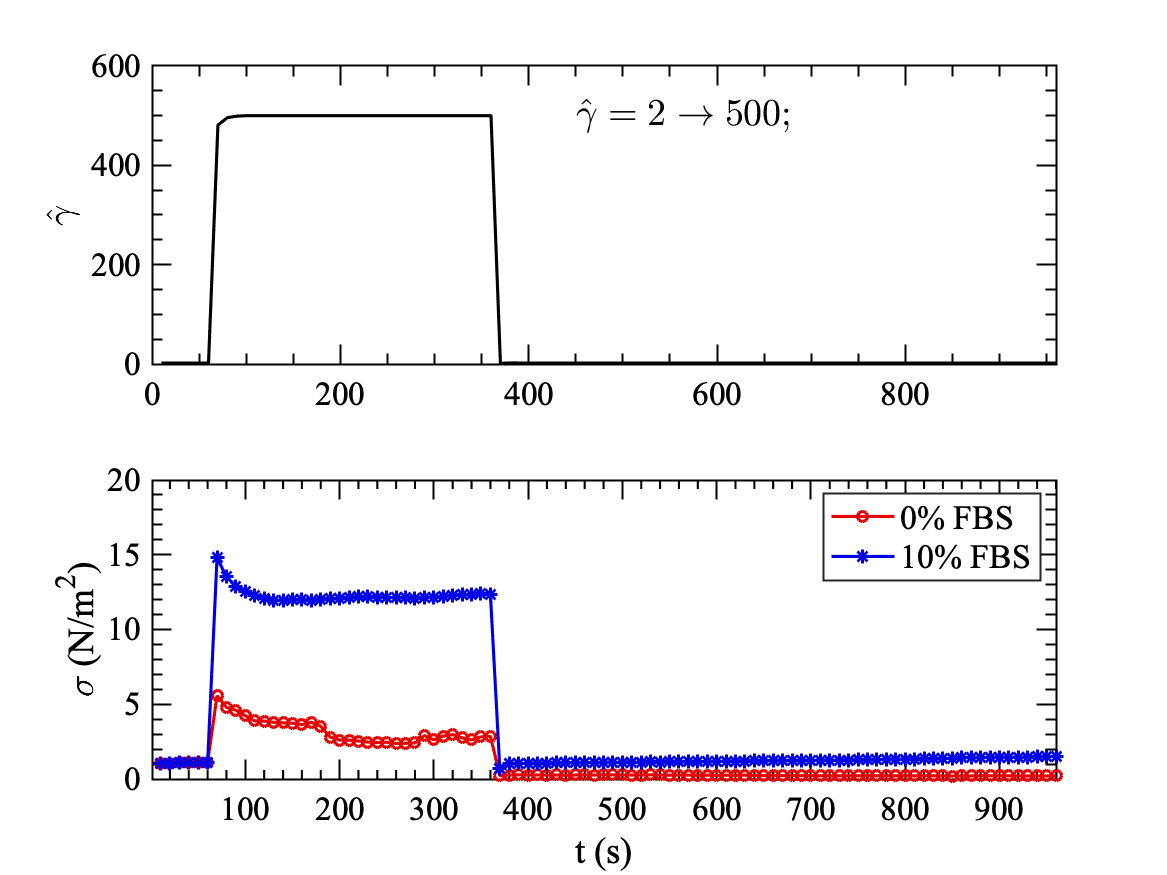}
		\caption{$\hat{\gamma}$ = 2 to 500}
		\label{fig:stress_5ss}
	\end{subfigure}
	\caption{\label{fig:stepStrain3}Comparison of $\sigma$ for Step Strain for 0\% FBS and 10\% FBS}
\end{figure}

   Healthy cells, with 10\% FBS, completely recover small deformation immediately as shown in Fig.~\ref{fig:stS02_08b}, confirming elastic behaviour of healthy monolayer cells
 in the linear viscoelastic regimes ($\hat{\gamma}$ = 2 to 8). In non-linear viscoelastic regime ($\hat{\gamma}$ = 2 to 500) as shown in Fig.~\ref{fig:sts02_5b}, we see almost full recovery with a
 small time lag. This may happen as the cell monolayer is deformed beyond the elastic limit, and deformation may reach a viscoplastic regime where the loss modulus $G^{\prime \prime}$ is larger than the storage 
 modulus $G^{\prime}$. In this large deformation limit, some link of cytoskeleton biopolymers might be broken permanently. However, after a long time, it is expected to regain the broken link, and
 material properties should return to the original value. Further confirming that healthy cell monolayers are flexible and capable of instant complete recovery of material properties in case of small
 deformation/strain and delayed recovery in large deformation (Fig.~\ref{fig:ssTD10fbs}). Whereas serum-free cell monolayer at 0\% FBS does not show complete recovery even at a small value of strain
 ($\hat{\gamma}$ = 2 to 8) as shown in Fig.~\ref{fig:stS02_08a}, confirming inability of starved cell monolayer for reconciliation in cell structure and cell monolayer morphology.
 The step strain of higher value ($\hat{\gamma}$ = 2 to 500) alter the rheological properties of serum-starved cell monolayer (Fig.~\ref{fig:sts02_5a}). The cell monolayer doesn't full
 recovery at 0\% FBS (Fig.~\ref{fig:ssTD0fbs}). This may happen as the crosslinking of the biomolecules inside the cells and at the cell-cell junction in the cell monolayer is once broken it will be
 unable to rebuild these crosslinks due to lack of serum. Hence rheological properties of starved cell monolayer at 0\% FBS after deformation reduces permanently.
 
 In Fig.~\ref{fig:stepStrain3}, the values of stress $\sigma$ are shown for the step strain experiments. In the linear viscoelastic regime ($\hat{\gamma}$ = 2 to 8) as shown in Fig.~\ref{fig:stress_08ss},
 the stress for starved cell monolayer increases instantly as the step deformation is applied and then decreases, whereas the stress in healthy cell monolayer increases with the step strain, but
 remain much lower than that of the starved cell monolayer, and then smoothly increases during the applied step strain. At the end of the step strain, the values of stress in cell monolayer with and without
 serum becomes equal. As the step strain is removed, the stress in healthy cell monolayer returns to initial stress before the step strain and remains constant with time, but the stress in the starved cell
 monolayer shows some variations with time after the step strain is removed. This observation shows that healthy cell monolayer can effectively distribute the stress due to small deformation and
 prevent the stresses from increasing abruptly in the cell monolayer. This may be because of the many dynamic protein filaments in the cell cytoskeleton responsible for rigidity and restructuring of the cells,
 redistributing the stress effectively in the cytoskeleton and throughout the cell monolayer. Whereas, in the starving cell monolayer, the cell cytoskeleton may not form these fibrous proteins and thus unable to
 distribute the stress effectively.
 
   For large deformations ($\hat{\gamma}$ = 2 to 500), as shown in Fig.~\ref{fig:stress_5ss}, the stress in a healthy cell monolayer is three times higher than for starved cell monolayer. The stress return to its
initial value after the step strain for cell monolayer with and without serum. The starved cell monolayer shows a slight decrease in stress with time after the step strain, whereas the stress remains
constant in the healthy cell monolayer. 
 
\section{Discussion and Conclusions}

   Serum starvation is standard procedure in cell biology \cite{Pirkmajer:2011fa}, but its effect on mechanical or rheological properties of the cells has not been extensively studied. \citet{Miyaoka:2011ej} reported the
micro-rheological properties of serum-starved 3T3 fibroblasts using AFM. They starved the cells at 0.1\% FBS for 24 hours and conducted the AFM study on single fibroblasts cells in a micro-fabricated glass
substrate. Whereas, in our study, we report the macro-rheological properties of cell monolayer with and without serum at 10\% and 0\% FBS, respectively. Our study has shown the significant difference in the
rheology of the cell monolayer for both the case a) with serum (10\% FBS) and b) without serum (0\% FBS) as well as at a) low shear strain and b) high shear strain. We can infer that the rheological properties
of cell monolayer depend on shear strain and concentration of serum in growth media. The experiments were conducted by placing the cell monolayer between two parallel plates.

  First, in the oscillatory shear experiments, we found that for low shear strain $(\hat{\gamma} \le 8)$, cell monolayer behaves like an elastic solid and both moduli, $G^{\prime}$ and $G^{\prime\prime}$,
remains constant for cell monolayer with and without-serum. As the strain is increased, interestingly both storage and loss moduli decrease, and after a certain strain value, loss modulus $(G^{\prime\prime})$
becomes larger than the storage modulus indicating the transition from an elastic solid-like to the viscous fluid-like behaviour of the cell monolayer. This crossover occurs at $\hat{\gamma} \sim 210$ for the
serum-starved cell monolayer (Fig.~\ref{fig:as1}) and $\hat{\gamma} \sim 450$  for healthy cell monolayer (Fig.~\ref{fig:as2}). This slightly delayed crossover of the healthy cell monolayer indicates that under
full serum conditions, cell monolayer stays elastic solid-like at higher strain values than that of the cell monolayer without serum. The slightly delayed transition from an elastic solid-like to the viscous
fluid-like behaviour in healthy cell monolayer with 10\% serum is attributed to the highly active or dynamic nature of the cytoskeleton of it's cells \cite{Stamenovic:2005kh, Trepat:2007jb}. Under full serum
conditions, cell cytoskeleton can restructure effectively in response to external deformations. However, without serum, many dynamic processes in the cell stop and the cytoskeleton become less active
\cite{Pirkmajer:2011fa}. Serum starvation reduces the ability of the cell to form new links in the structure of cytoskeleton, and when these links are broken due to deformation of the cells, they are not
rebuilding as effectively as a healthy cell monolayer with 10\% FBS. This results in the early transition of serum-starved cell monolayer to fluid-like behaviour. Also, serum-starved cell monolayer cannot
recover from even the small deformation (Fig.~\ref{fig:ssTD0fbs}), whereas, a healthy cell monolayer with full serum can recover from even large deformations (Fig.~\ref{fig:ssTD10fbs}). Our values of storage
and loss moduli are consistent with the data of \citet{Dakhil:2016ij} on HeLa cells at the gap of 15 $\mu$m.

   Our results show the strain-softening of the cell monolayer similar to many other studies on cells \cite{Trepat:2007jb, Krishnan:2009kk}, but some other studies on cells reported strain-stiffening
\cite{Pourati:1998we, Wang:2002bu, Fernandez:2006kt}. This paradox of strain stiffening and softening was studied by \citet{Wolff:2012bx} on minimal cytoskeleton model system (F-actin/HMM) which shows
similar softening-stiffening behaviour. They proposed that the stiffening is a direct viscoelastic response to applied stress and caused by the non-linear stretch resistance of individual semiflexible
biopolymers, whereas, the softening is characterized as inelastic fluidization which is caused by the dynamical evolution of the mutual bonds between the biopolymers due to the applied strain. In our work,
we believe, that the softening emerges as the characteristic response to the applied strain because of the evolution of the bonds between the biopolymers and cell-cell contacts in the monolayer. Our results
also shows the soft glassy behaviour of the cell monolayers (Fig.~\ref{fig:ampSND}). The soft glassy nature for the healthy cell monolayer with 10\% FBS (Fig.~\ref{fig:ndas2}) is more pronounced
than for the starved cell monolayer (Fig.~\ref{fig:ndas1}), again confirming that in the absence of the serum, the cell monolayer looses the ability to restructure itself.

   In frequency sweep experiments, our findings suggest that for small deformation in the linear viscoelastic regime, the cell monolayer with and without serum shows weak power-law dependence on the 
applied frequency. Our observation of macro-rheological storage modulus $(G^{\prime})$ and loss modulus $(G^{\prime\prime})$ of the cell monolayer without serum at 0\%FBS for the small deformation
($\hat{\gamma}$ = 5) in linear viscoelastic regime shows a weak power-law dependence with an exponent of 0.15 and 0.14 throughout the applied frequencies range, respectively. However, for healthy cell monolayer
with 10\% FBS, only the storage modulus $(G^{\prime})$ shows the weak power-law dependence (exponent = 0.13) throughout the applied frequency range. Whereas, the loss modulus  $(G^{\prime\prime})$
shows weak dependence (exponent = 0.13) for low frequencies ($\omega \le 20$rad/s) and a stronger power-law dependence with an exponent of 0.27 at higher frequencies ($\omega \ge 20$rad/s). 

   In the non-linear viscoelastic regime, at high deformations with $\hat{\gamma}$ = 500 and 5000, serums starved cell monolayer shows a strong power-law dependency for high frequency 
($\omega \ge 20$rad/s (Fig.~\ref{fig:frS0})). This indicates that serum-starved monolayer becomes stiff at the large values of strains at high frequency. Whereas, the cell monolayer with full serum at 10\% FBS
remains flexible even at the large values of strain and at high frequencies, as power-law dependency remains weak for the cell monolayer with 10\% FBS (Fig.~\ref{fig:frS10}). \citet{Miyaoka:2011ej} also
reported that for starved NIH3T3 cells (0.1\% FBS for 24 hrs) the local storage modulus $(G^{\prime})$ inside the cytoskeleton of NIH 3T3 cells shows the similar dependence with frequency. They reported a
weak power-law dependence at lower frequencies, and at higher frequencies, the power-law exponent increases. We can infer that, under serum starvation, macro and micro storage modulus tend to
show a similar frequency dependence. 

 Next, we performed the step strain experiments to study the recovery of applied deformation to the cell monolayer. We applied steps of (a) $\hat{\gamma}$ = 5 in linear viscoelastic regime and (b)
$\hat{\gamma}$ = 500 in non-linear viscoelastic regime. Our study reveals that a healthy cell monolayer with 10\% FBS shows a complete recovery of storage and loss moduli even at large deformation
(step strain from $\hat{\gamma}$ = 2 to 500, (Fig.~\ref{fig:sts02_5b})), whereas a starving cell monolayer doesn't recover completely even for small deformations in linear viscoelastic regime
(step strain from $\hat{\gamma}$ = 2 to 8, (Fig.~\ref{fig:stS02_08a})). Also, the stress in healthy cell monolayer increases with very low gradient during the small deformation, but the stress in the
serum-free cell monolayer shoots up instantly at a much higher value and then decreases with steep gradient during the step strain Fig.~\ref{fig:stress_08ss}. For high step strain with $\hat{\gamma}$ = 500, the
stress in a healthy cell monolayer is higher than that of the serum-free cell monolayer. In both cases, stress decreases during the applied step strain and recovers instantly when the step strain was removed. The
stress shows a slight decrease in values for the serum-free cell monolayer after the step strain. This observation shows that healthy cell monolayer can distribute stress more effectively than that
of the serum-free or starved cell monolayer. This may be because of the many dynamic protein filaments in the cell cytoskeleton such as actomyosin network, microtubules, intermediate filaments
responsible for rigidity and restructuring of the cells, redistributing the stress effectively in the cytoskeleton and throughout the cell monolayer. Whereas, in the starving cell monolayer, the cell cytoskeleton may
not form these fibrous proteins and cannot distribute the stress effectively.

 \citet{Harris:2012hn} has shown that the monolayer mechanical properties are strongly dependent on the actin cytoskeleton, myosin, and intercellular adhesions interfacing adjacent cells. In their study they
 harvested the cell monolayer and did the stretching experiments to access the mechanical properties of the monolayer. Their study revealed the role of keratin filaments in cell monolayer mechanics. In another 
 study by \citet{Helfand:2011dx}, it is shown that serum starvation has significant effect on vimentin intermediate filaments organization and strongly affect the lamellipodia formation in 3T3 fibroblasts. 
    
       Our study of the bulk rheology of the cell monolayers reveals that the individual cells cytoskeleton, intermediate filaments, and intercellular adhesions play a defining role in the rheological properties of the
cell monolayer. The cytoskeleton, comprising of various biopolymers, and active proteins, can restructure itself by making and breaking filamentous biopolymers. The cell cytoskeleton helps the cells
and the cell monolayer as a whole to reorient and restructure itself in the direction of applied strain. Serum starvation leads to reduced protein synthesis and hence reduced cell growth 
\cite{Pirkmajer:2011fa, Helfand:2011dx}. This lead to less active cells with the less dynamic cytoskeleton of the cells. In the absence of serum, which contains all the nutrients and growth factors, cells
become less active, and its cytoskeleton does not effectively restructure itself in response to the applied strain. In our experiments, we have shown that the cell monolayer, under serum-free conditions, appears
to be softer than the monolayer which is supplied with all the nutrients and growth factor, i.e. with 10\% FBS. In the presence of serum, cells are healthy and more dynamic or active, and can effectively
restructure its cytoskeleton under given strain or stress. This leads to its increased strength and more dynamic behaviour.

\section*{Conflicts of interest}
There are no conflicts to declare.

\section*{Acknowledgements}
We acknowledge J. W. Nelson for sending the MDCK cell line. The authors acknowledge the support from the Science and Engineering Research Board, Department of Science \& Technology,
(SERB-DST) grant (no. CRG/2020/003342 ) funded by Government of India. The authors also would like to acknowledge Indian Institute of Technology, Ropar for laboratory usage and financial support.





\bibliography{BioRheo1a} %

\end{document}


%
%
%
%
%


\section{\label{sec:intro}Supplementary Information}

\subsection{Experimental Setup and Protocols}

\begin{figure}[htbp]
	\begin{center}
		\includegraphics[width=0.6\columnwidth]{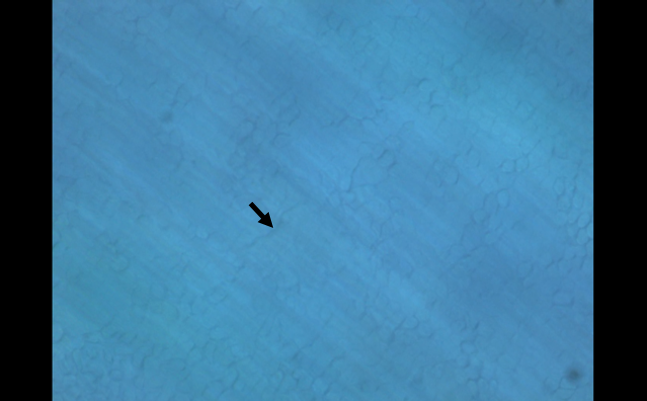}
		\caption{\label{fig:CM_shear}Cell Monolayer during Oscillatory Shearing Experiment}
	\end{center}
\end{figure}
Fig.~\ref{fig:CM_shear}: The arrow shows the cell monolayer which is showing the movement of cell monolayer during the shearing experiment in the accompanying video with this SI.

\begin{figure}[htbp]
\begin{center}
\includegraphics[width=0.6\columnwidth]{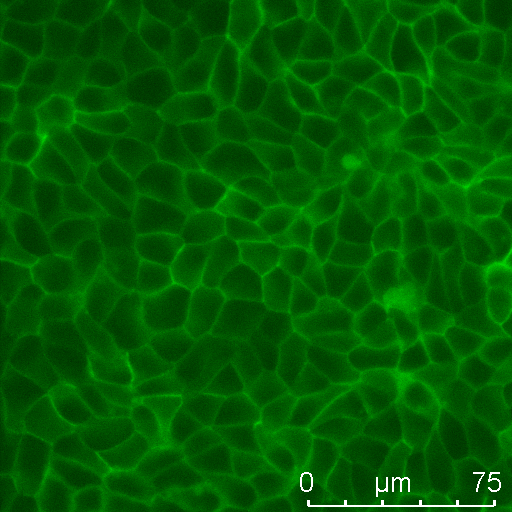}
\caption{Fluorescence image of MDCK GFP-E-Cadherin cell monolayer before experiment}
\label{fig:MDCK1}
\end{center}
\end{figure}

Fig.~\ref{fig:MDCK1}: Fluorescence image of MDCK II cells. These cells are stably transfected with GFP-E-cadherin.

\begin{figure}[!tbp]
	\begin{subfigure}[b]{0.45\textwidth}
		\includegraphics[width=\textwidth]{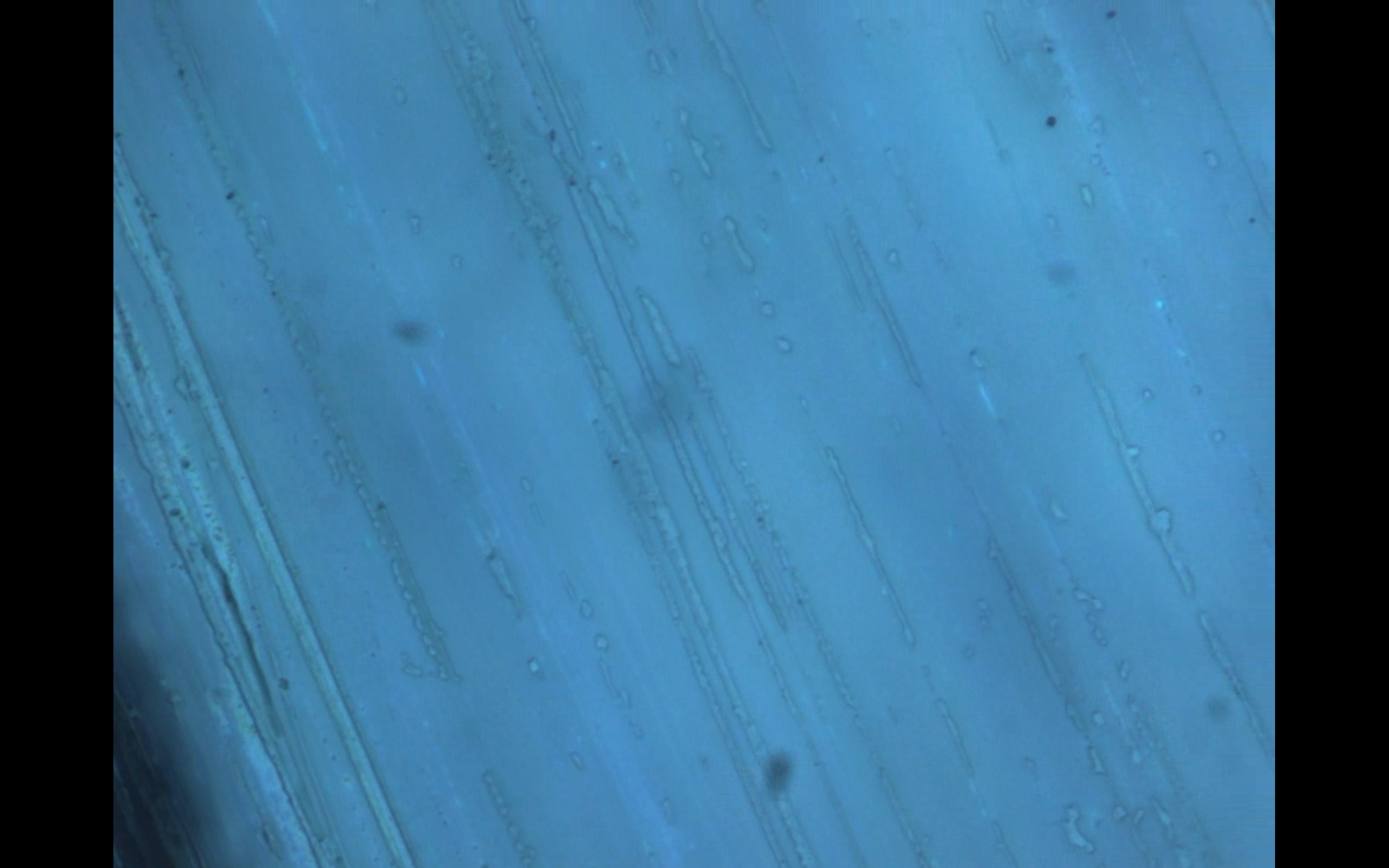}
		\caption{}
		\label{fig:bgp1}
	\end{subfigure}
	\hfill
	\begin{subfigure}[b]{0.45\textwidth}
		\includegraphics[width=\textwidth]{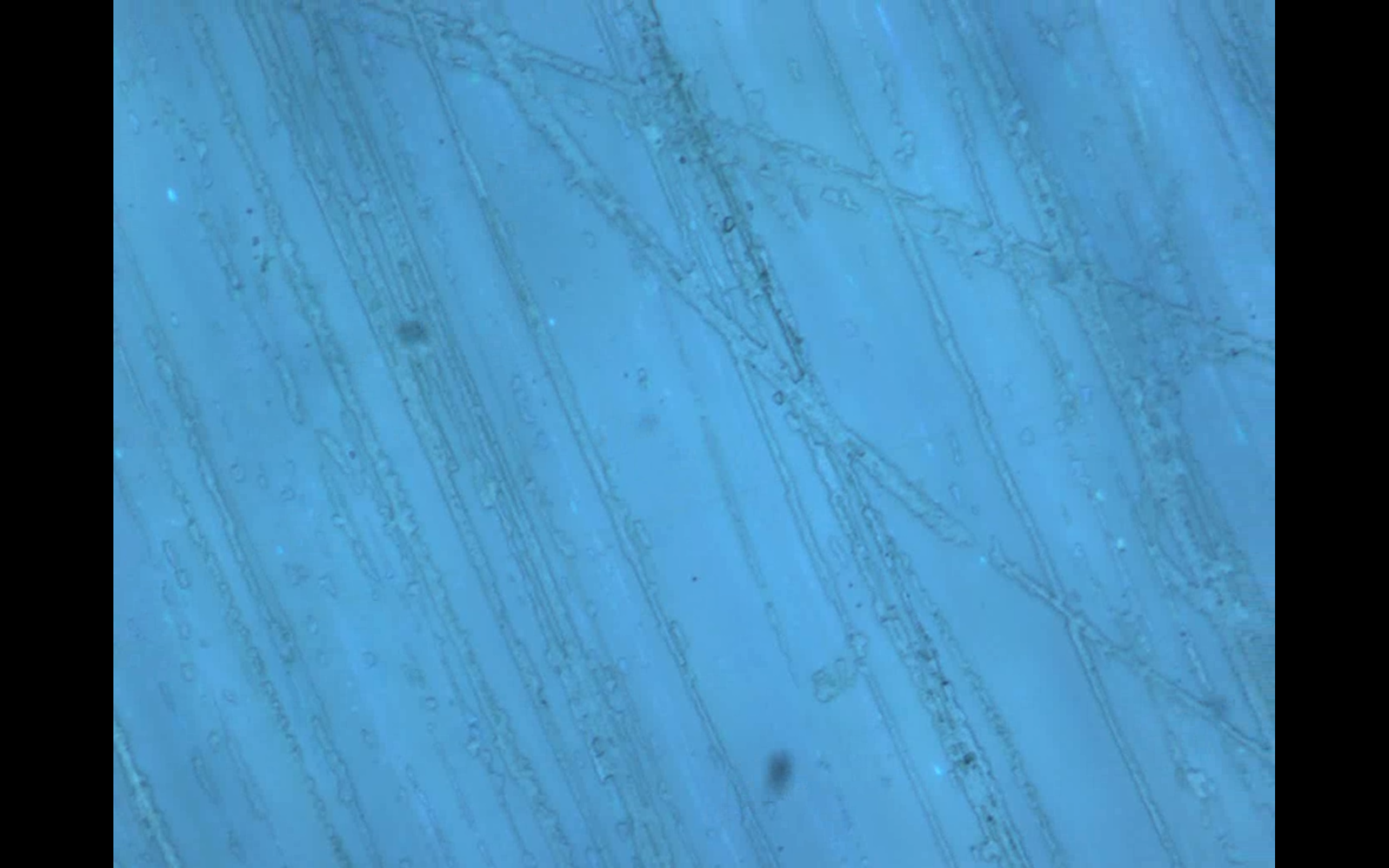}
		\caption{}
		\label{fig:bgp2}
	\end{subfigure}
	\hfill
	\begin{subfigure}[c]{0.45\textwidth}
		\includegraphics[width=\textwidth]{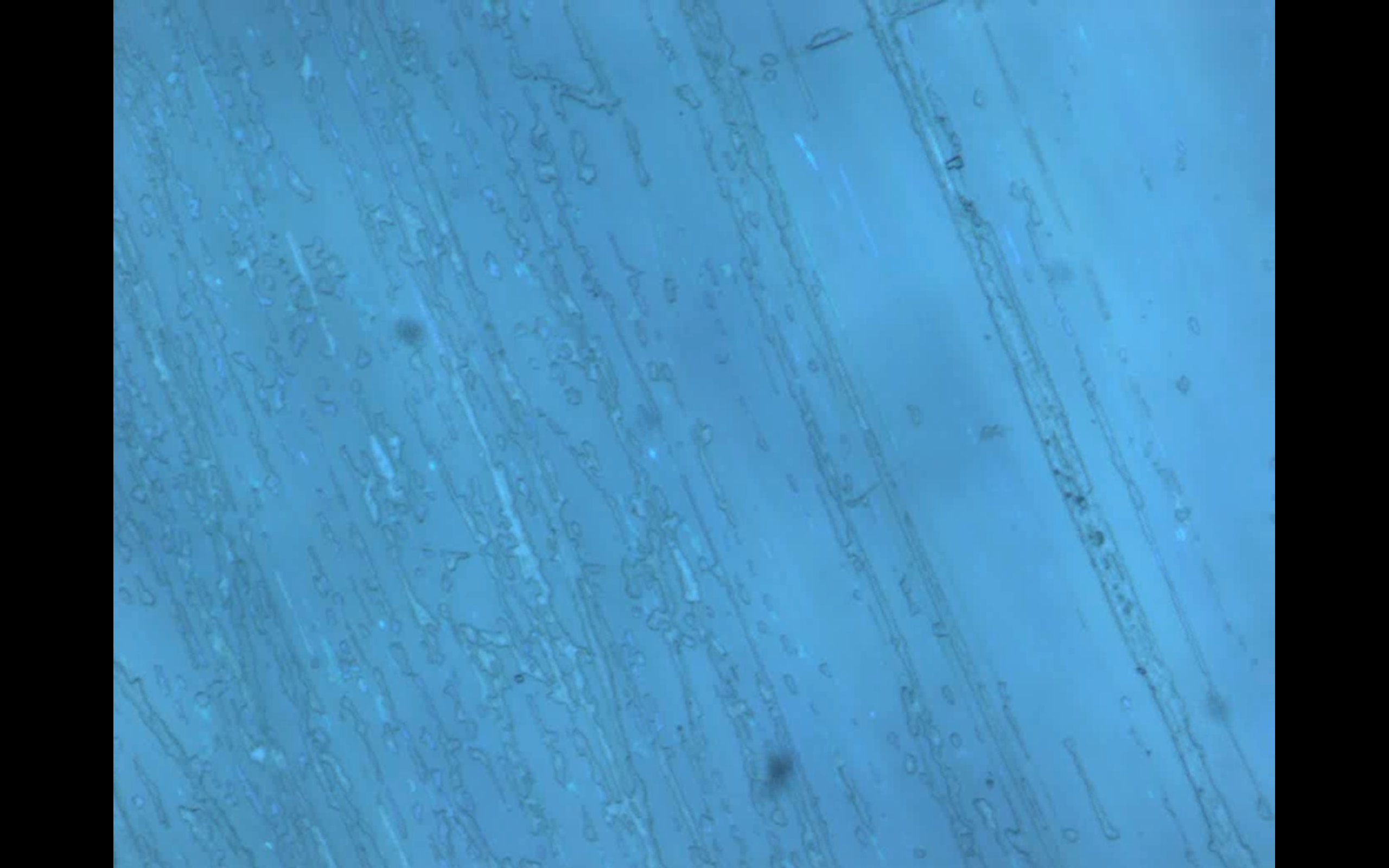}
		\caption{}
		\label{fig:bgp3}
	\end{subfigure}
	\caption{\label{fig:bgp}Focusing of bottom glass plate: (a) Bottom Glass Plate (outer part at the edge), (b) Bottom Glass Plate (outer part far from edge) (c) Bottom Glass Plate (central part).
	Focused using 20X microscope of the microscope module of the rheometer.}
\end{figure}

 Fig.~\ref{fig:bgp} shows the focused images of the bottom glass plate of microscope module of the rheometer. We focused on the upper surface of bottom glass plate using 20X objective on the microscope module
of the rheometer which is seen in Fig.~\ref{fig:bgp}. Without changing the focus ( z- axis), we track the surface from outer side of the glass bottom plate towards the centre. Fig.~\ref{fig:bgp1} is the focused image
of the surface at the periphery or outer side of the glass plate.  Fig.~\ref{fig:bgp2} is the images at the outer side far from the edges and Fig.~\ref{fig:bgp3} is at the centre of the bottom glass plate.
 
\begin{figure}[!tbp]
	\begin{subfigure}[b]{0.45\textwidth}
		\includegraphics[width=\textwidth]{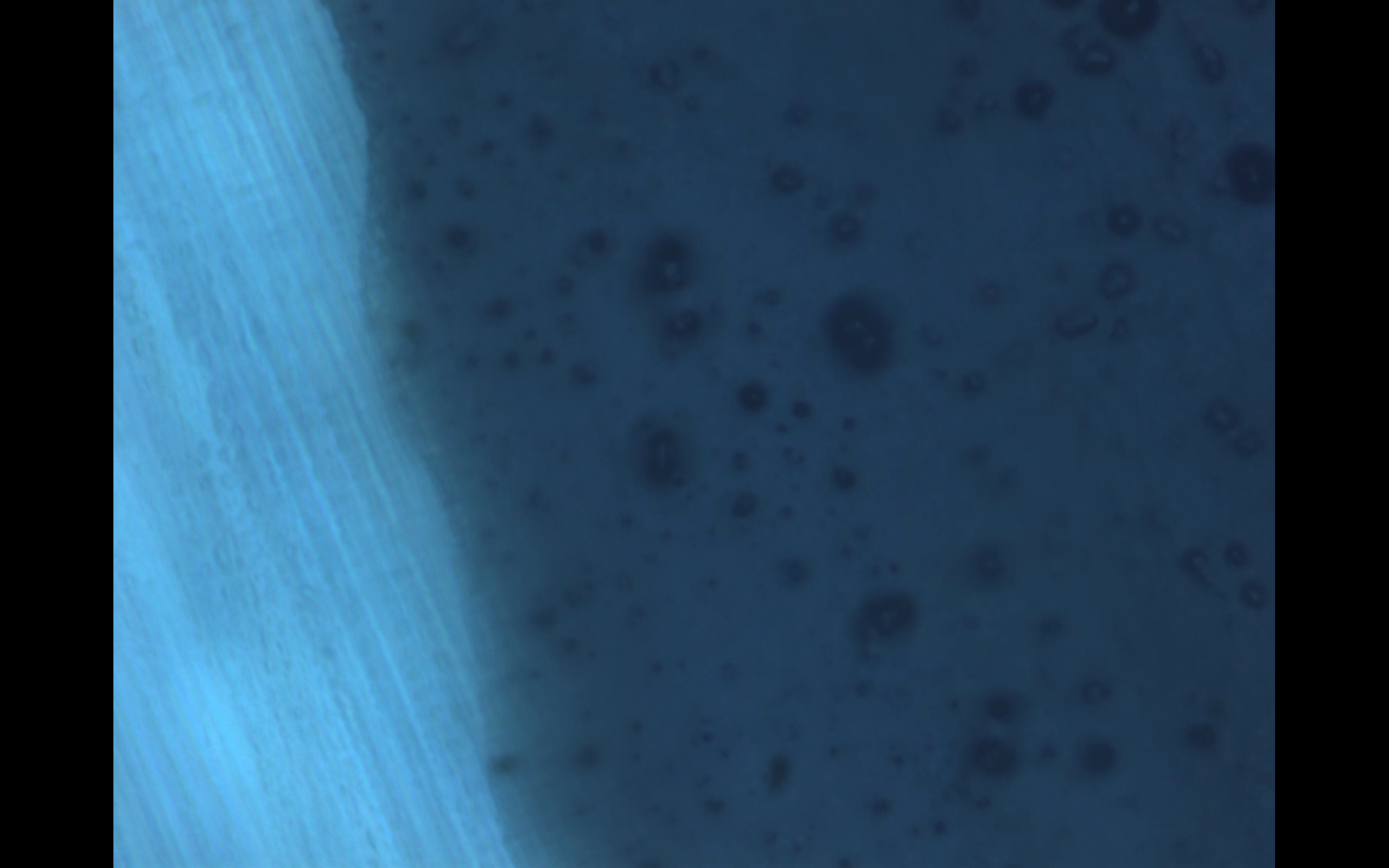}
		\caption{}
		\label{fig:ugp1}
	\end{subfigure}
	\hfill
	\begin{subfigure}[b]{0.45\textwidth}
		\includegraphics[width=\textwidth]{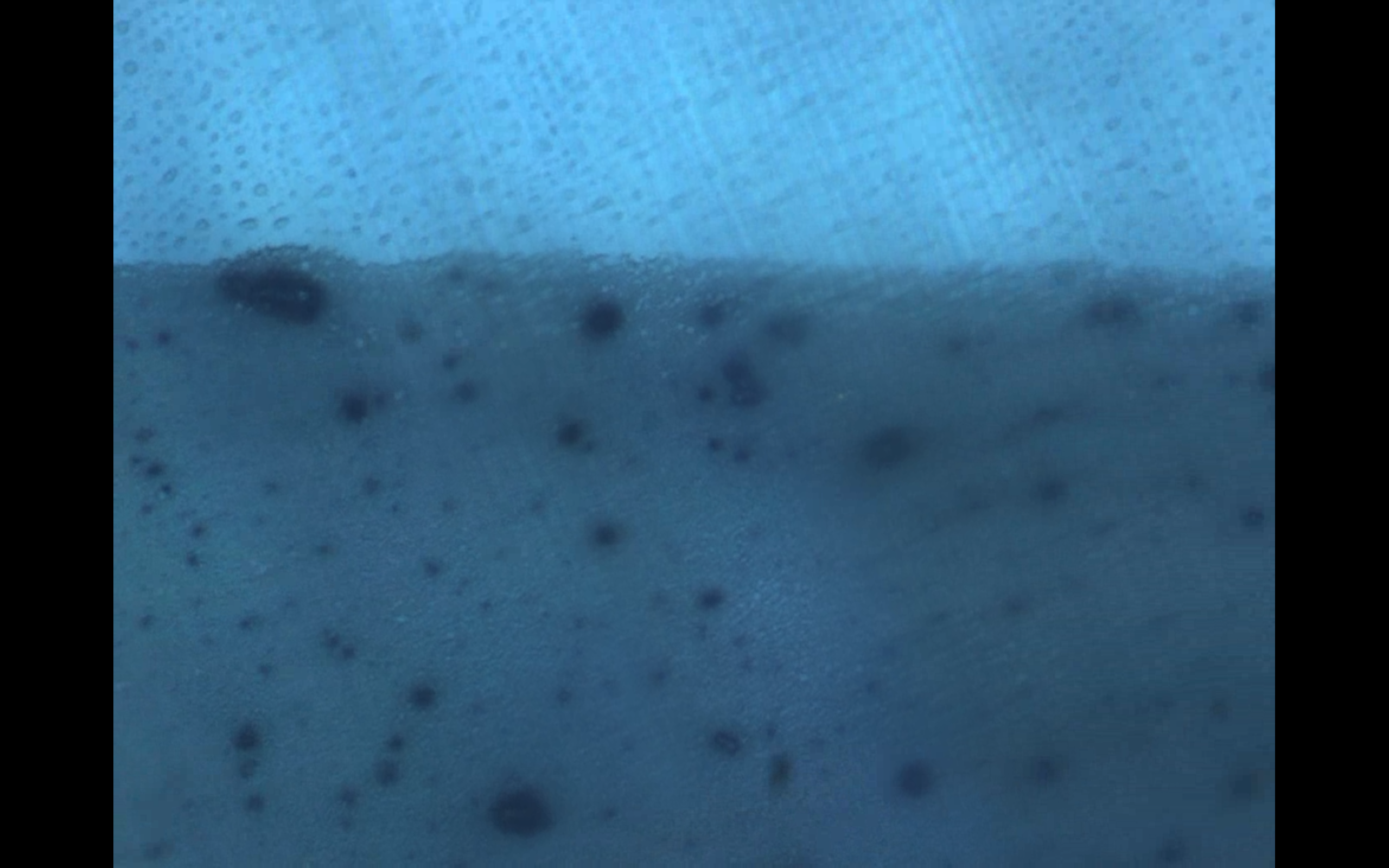}
		\caption{}
		\label{fig:ugp2}
	\end{subfigure}
	\hfill
	\begin{subfigure}[c]{0.45\textwidth}
		\includegraphics[width=\textwidth]{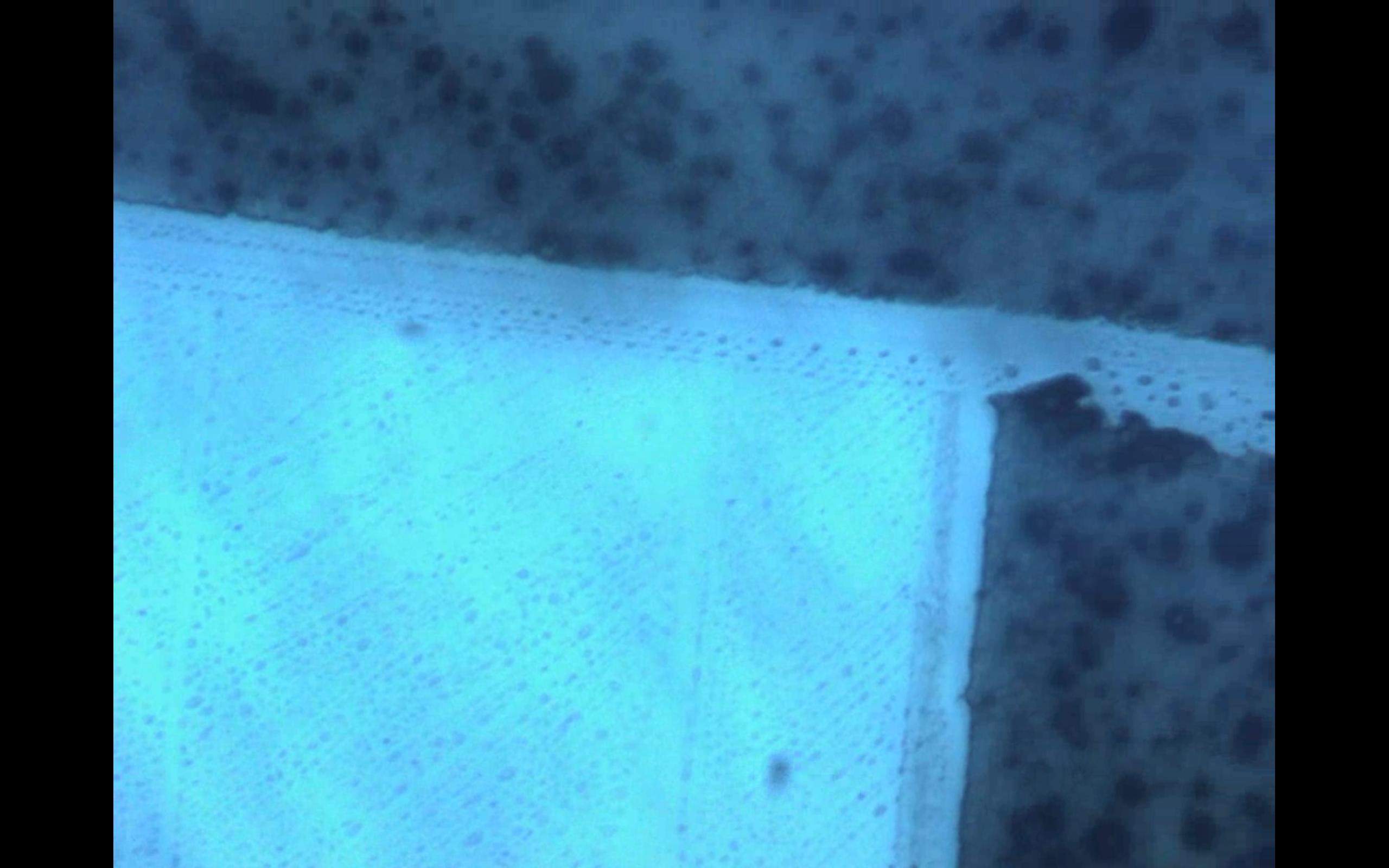}
		\caption{}
		\label{fig:ugp3}
	\end{subfigure}
	\caption{\label{fig:ugp}Focusing of upper glass coverslip: (a) Upper Glass Coverslip (outer part at the edge), (b) Upper Glass Coverslip (outer part far from edge) (c) Upper Glass Coverslip (central part).
	Focused using 20X microscope of the microscope module of the rheometer.}
\end{figure}

 Fig.~\ref{fig:ugp} shows the focused images of the upper glass coverslip which is attached to the upper steel probe. To focus on the upper glass coverslip, we make lines on the glass coverslip using a marker. Then
we focus in this edge of this marker and track the edges across the glass coverslip using 20X objective on microscope module of the rheometer. Without changing the focus (z- axis), we track the edges from
outer side of the glass coverslip towards the centre. Fig.~\ref{fig:ugp1} is the focused image of the mark's edge at the periphery or outer side of the glass coverslip.  Fig.~\ref{fig:ugp2} is the images at the outer
side far from the edges and Fig.~\ref{fig:ugp3} is at the centre of the upper glass coverslip.

Protocol to perform thin gap rheology of cell monolayer:

\begin{enumerate}
\item We set the zero gap on bottom glass plate of 65 mm diameter of the microscope module (Rheo-Microscope) of the rheometer. For this we use the 25mm steel upper plate which is already glued with the
25mm circular glass coverslip and the glass bottom plate. 
\item We fix the rectangular glass coverslip on which we have already grown a confluent circular cell monolayer
\item We lower the upper probe/plate to the gap width of 300 $\mu$m.
\item We then use the microscope to get the live video of cell monolayer. And keep lowering the upper plate to reduce the gap.
\item We also monitor the normal force sensed by the upper plate. It shows 0 until the upper probe touches the material (cell monolayer).
\item As soon as the upper probe touches the cell monolayer the normal force increases.
\item We set the gap when the normal force is almost constant at ~0.1 N.
\item Also through the microscope, we can observe the slight change in shapes of the cells in the monolayer as it is squeezed by the upper plate.
\item At this gap height, we are ready to perform our experiments.
\end{enumerate}

\subsection{Videos}

\begin{itemize}
\item \verb|SI_mov1_CellMonolayer_Shearing.mp4| -- This video shows the cell monolayer as it is being sheared during the oscillatory shearing experiment. The video was captured using the microscopic module
of the MCR-702 Rheometer from Anton-Paar at 20X.  \\

%
\end{itemize}